\documentclass[12pt,epsf]{article}
\usepackage{amsmath,amsfonts,amssymb}
\usepackage{graphicx}
\usepackage{enumerate}

\newcommand{\bbibitem}[1]{\bibitem{#1}\marginpar{#1}}

\newcommand{\figref}[1]{Fig. \ref{#1}}
\newcommand{\secref}[1]{Sec. \ref{#1}}
\newcommand{\tableref}[1]{Table \ref{#1}}
\newcommand{\appref}[1]{Appendix \ref{#1}}

\def\Label#1{\label{#1}%
  \smash{\hbox to0pt{\raise1ex\hbox{\tiny[#1]}\hss}}}
\def\noLabels{\let\Label=\label}
\def\nobbibitem{\let\bbibitem=\bibitem}

\newcommand{\be}{\begin{equation}}
\newcommand{\ee}{\end{equation}}
\newcommand{\bea}{\begin{eqnarray}}
\newcommand{\eea}{\end{eqnarray}}
\newcommand{\beas}{\begin{eqnarray*}}
\newcommand{\eeas}{\end{eqnarray*}}
\newcommand{\ba}{\begin{array}}
\newcommand{\ea}{\end{array}}

\newcommand{\lt}{\left}
\newcommand{\rt}{\right}

\makeatletter
\@addtoreset{equation}{section}
\makeatother

\renewcommand{\d}{\partial}

\newcommand{\tr}{\mathrm{Tr}}

\newcommand{\nbox}{{\,\lower0.9pt\vbox{\hrule \hbox{\vrule height 0.2 cm \hskip 0.19 cm \vrule height 0.2 cm}\hrule}\,}}

\def\href#1#2{#2}

\begin{document}
\begin{titlepage}
\hfill
\vbox{
    \halign{#\hfil         \cr
           } 
      }  
\vspace*{20mm}
\begin{center}
{\Large \bf  Twisted Inflation}

\vspace*{15mm}
\vspace*{1mm}
{Joshua L. Davis,\footnote{\tt e-mail: jdavis@phas.ubc.ca} Thomas S. Levi,\footnote{\tt e-mail: tslevi@phas.ubc.ca} Mark Van Raamsdonk,\footnote{\tt e-mail: mav@phas.ubc.ca} and Kevin R.L. Whyte\footnote{\tt e-mail: kwhyte@phas.ubc.ca}}
\vspace*{1cm}

{\it Department of Physics and Astronomy,
University of British Columbia\\
6224 Agricultural Road,
Vancouver, B.C., V6T 1W9, Canada}

\vspace*{1cm}
\end{center}

\begin{abstract}

We present a new mechanism for slow-roll inflation based on higher dimensional supersymmetric gauge theory compactified to four dimensions with twisted (supersymmetry breaking) boundary conditions. These boundary conditions lead to a potential for directions in field space that would have been flat were supersymmetry preserved. For field values in these directions much larger than the supersymmetry-breaking scale, the flatness of the potential is nearly restored. Starting in this nearly flat region, inflation can occur as the theory relaxes towards the origin of field space. Near the origin, the potential becomes steep and the theory quickly descends to a confining gauge theory in which the inflaton does not exist as a particle. This confining gauge theory could be part of the  Standard Model (QCD) or a natural dark matter sector; we comment on various scenarios for reheating.  As a specific illustration of this mechanism, we discuss 4+1 dimensional maximally supersymmetric gauge theory on a circle with antiperiodic boundary conditions for fermions. When the theory is weakly coupled at the compactification scale, we calculate the inflaton potential directly in field theory by integrating out the heavy W-bosons and their superpartners. At strong coupling the model can be studied using a gravity dual, which realizes a new model of brane inflation on a non-supersymmetric throat geometry. Assuming there exists a UV completion that avoids the $\eta$-problem, predictions from our model are consistent with present observations, and imply a small tensor-to-scalar ratio.

\end{abstract}

\setcounter{footnote}{0}
\end{titlepage}

\vskip 1cm
\section{Introduction}

Cosmic inflation \cite{Guth:1980zm,Linde:1981mu,Albrecht:1982wi} provides an attractive explanation for the flatness and homogeneity of the observable universe and gives rise in a natural way to primordial inhomogeneities that lead to structure formation (for a recent review, see \cite{Baumann:2009ds}). Many effective field theory models of inflation have been suggested, however, one would like to have some guiding principles that narrow the field and therefore allow more specific predictions. From the field theory point of view, an attractive condition to demand is naturalness: the parameters in the effective field theory should not require a significant amount of fine-tuning. Another possible set of constraints may arise by demanding that the theory has a consistent UV completion including gravity. One possibility for such a UV completion is string theory; if we assume that string theory gives the correct microscopic theory of our universe, then the correct model of inflation must arise in some way from string-theoretic degrees of freedom.

\begin{figure}[t]
\begin{center}
\includegraphics[width=0.7\textwidth]{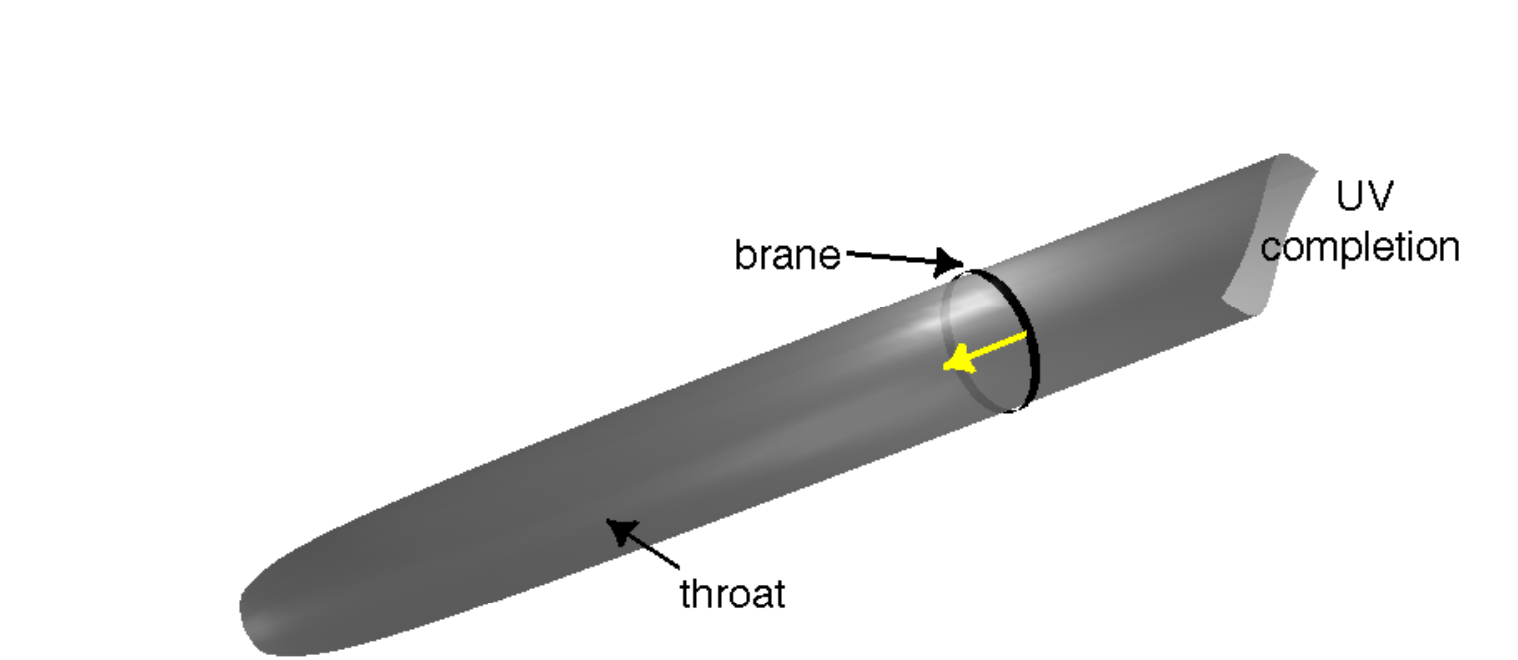}
\end{center}
\caption{\label{fig-throat} Geometrical picture of inflation in the strongly coupled regime.}
\end{figure}

In this paper, we consider a new scenario for inflation that can be naturally embedded in string theory, but more generally can be understood as a mechanism to generate a naturally flat inflaton potential from supersymmetry-breaking compactification of higher-dimensional supersymmetric gauge theory. The model has a strongly-coupled version best described as inflation coming from wrapped branes on a non-supersymmetric throat geometry, and a weakly-coupled version that can be analyzed directly in field theory. For any coupling, the model has a natural embedding in string theory based on the low energy dynamics of D-branes or M5-branes. We begin by describing the basic idea in a general context.
\vskip 0.1 in
\noindent
{\bf The brane inflation scenario}
\vskip 0.1 in
\noindent
Suppose that as part of some warped compactification of string theory, we have a supersymmetric throat geometry which includes a cycle upon which a brane is wrapped (it is also extended in the four non-compact directions). This is schematically depicted in \figref{fig-throat}. Suppose further that there is a
 moduli space for the brane, so that it can move freely up and down the
 throat. Now, consider a related geometry for which we choose some different
 supersymmetry-breaking boundary conditions on this cycle, such that
 in the modified non-supersymmetric geometry, the cycle contracts to zero
 size at the tip of the throat. Far away from this tip, where the
 cycle is large, the geometry is very similar to the supersymmetric case, so
 the radial potential for the brane is almost flat. But eventually, the
 brane will roll down to the tip and disappear, giving all its energy to the
 other degrees of freedom in the theory when it contracts. Thus, the scalar field describing the radial position of the brane is a natural candidate for an inflaton field.
\vskip 0.1 in
\noindent
{\bf The field theory scenario}
\vskip 0.1 in
\noindent
Consider a higher-dimensional supersymmetric field theory with a moduli space (i.e. flat directions in the potential) parameterized by some scalar fields. Now, compactify to 3+1 dimensions on a manifold for which supersymmetry is broken by boundary conditions (e.g. a circle with antiperiodic boundary conditions for fermions). Suppose that the compactified theory flows in the IR to a massive (confining) theory.

In the uncompactified theory, if we take one of the moduli scalar fields $\phi$ to have a large expectation value, then typically there will be some W-bosons (and their superpartners) that become very massive. With enough supersymmetry, integrating these out does not spoil the existence of a moduli space, so the contribution to the effective potential coming from the massive bosons must exactly cancel the contribution from the fermions. This remains true in a supersymmetric compactification but is generically modified by boundary conditions that break supersymmetry.

To give a concrete example, consider  a ``twisted" circle compactification with periodic boundary conditions  for the bosons and  antiperiodic boundary conditions for the fermions on circle. The mode numbers for the Kaluza-Klein (KK) modes of the W-bosons are shifted relative to their fermionic superpartners, so the masses are
\bea
M_b &=& \sqrt{\phi^2 + {n^2 \over R^2}} ~, \cr
M_f &=& \sqrt{\phi^2 + {(n+ {1 \over 2})^2 \over R^2}}~,
\eea
where $1/R = M_{KK}$ is the KK compactification scale.  Thus, there is no longer an exact cancelation in calculating the effective potential. However, boson and fermion modes with $n \ll \phi R$ will be almost degenerate. As a result, cancelations return for large $\phi R$ and the potential becomes flat here.

Thus, scalar fields which would be moduli in a supersymmetric compactification
give rise to fields that can act as inflaton fields in the twisted compactification. If these fields start with large values $\phi \gg 1/R$, they will roll slowly back towards the origin of moduli space. At the origin, we get to the theory which is massive in the IR, so there are no more light excitations.
\vskip 0.1 in
\noindent
{\bf A specific model}
\vskip 0.1 in
\noindent
In the bulk of this paper, we will illustrate these two scenarios in a specific model. We will see that the brane inflation picture and the field theory picture represent strongly coupled and weakly coupled versions of the same mechanism.

The model we consider is maximally supersymmetric 4+1 dimensional $U(N)$ Yang-Mills theory compactified on a spatial circle of radius $R = 1/M_{KK}$ with antiperiodic boundary conditions for fermions \cite{Witten:1998zw}.\footnote{In the UV, this can be completed as the 5+1 dimensional $(0,2)$ CFT (the low-energy theory of M5-branes) compactified on a torus with periodic boundary conditions on one cycle and antiperiodic boundary conditions on the other cycle.}  The effective dimensionless 't Hooft coupling of the compactified theory is $\lambda$.

We consider initial conditions in which one of the adjoint scalar fields is $X = diag(\phi,0,\cdots,0)$. With supersymmetry-preserving boundary conditions, the scalar  $\phi$ would parameterize an exact flat direction in the potential. In the case of broken supersymmetry, a potential is generated in the quantum effective action, but this turns out to be quite flat for $\phi \gg M_{KK}$ . When the theory is weakly coupled at the Kaluza-Klein scale ($\lambda \ll 1$), we can compute the effective potential by integrating out the W-bosons and their superpartners at one loop. We find that the potential for large $\phi$ (canonically normalized) takes the form
\be
V(\phi) = V_0 \left(1- A (\phi / \phi_0)^2 e^{-{\phi \over \phi_0}} \right) \; ,
\ee
which quickly approaches a constant for large $\phi$.

At strong coupling ($\lambda \gg 1$), the field theory has a gravity dual description as type IIA string theory on a non-supersymmetric D4-brane throat geometry or M-theory on a non-supersymmetric M5-brane throat at very strong coupling.\footnote{We will make these statements more precise in the body of the paper.} In either case, the circle direction corresponding to the antiperiodic compactified direction of the field theory contracts smoothly to a point at the IR end of the geometry. Assuming that $N$ is large, taking one scalar to have an expectation value corresponds to introducing a probe brane in this throat geometry which is extended along all the field theory directions, including the compact ones, as shown in figure \figref{fig-throat}. This is the wrapped brane that we described in general above. The quantum effective potential for our scalar is then calculated using the effective action for this probe brane (Born-Infeld plus Chern-Simons type couplings to the form fields) in the throat geometry. In either the type IIA string theory regime or the M-theory regime, we find a potential that behaves for large $\phi$ as
\be
V(\phi) = V_0(1-  \phi_0^3/\phi^3) \; ,
\ee
again asymptotically approaching a constant.

Starting with the effective actions we derive and {\it minimally coupling} to gravity,\footnote{A matter field is minimally coupled to gravity if its action does not involve explicit curvature tensors.} it is straightforward to show that we can achieve slow-roll inflation with an acceptable spectrum of fluctuations for natural values of the parameters. For example, the predictions for the scalar spectral index $n_s$ range continuously from $n_s \approx 0.967$ in the weakly coupled model to $n_s \approx 0.973$ at strong coupling. The predictions and constraints for the minimally coupled model are summarized in \tableref{table:summary}.

\begin{table}[t]
\centering
\vskip 0.1 in
\begin{tabular}{|  c | c | }
\hline
$\lambda \ll 1  $ & $\lambda \gg 1$  \\   \hline   \, & \\

 $\qquad n_s = 1 - {2 \over {\cal N}_{CMB}} \sim 0.967 \qquad$  &  $\qquad n_s = 1 - {8 \over 5 {\cal N}_{CMB}} \sim 0.973 \qquad$ \\
& \\
 $\alpha_s = - {2 \over {\cal N}^2_{CMB}} \sim -0.00055$  &  $\alpha_s = - {8 \over 5 {\cal N}^2_{CMB}} \sim -0.00044$ \\
& \\
$r  \sim 8.6 \times 10^{-12}  N^{-1}$   &  $r  \sim 2.5 \times 10^{-9}  (\lambda N)^{3 \over 7}$ \\
\hline
\end{tabular}
\caption{Weak and strong coupling predictions for the scalar spectral index $n_s$, the running of the spectral index $\alpha_s$, and the tensor/scalar ratio $r$. For $\lambda \gg 1$ we require $N \gg 1$ for the analysis to be reliable, and $N \lesssim 10^4 \lambda^{-{3 \over 5}}$ to obtain enough e-foldings. Numerical values are given assuming CMB perturbations at the pivot scale left the horizon at ${\cal N}_{CMB} = 60$ e-foldings before the end of inflation. }
\label{table:summary}
\end{table}

On the other hand, it is well known that once the field theory is coupled to gravity and treated in an inflating spacetime, there will generically be $H^2 \phi^2$ terms in the effective potential (here $H$ is the Hubble parameter during inflation) that can spoil slow-roll inflation. The coefficient of such a term will in general depend on the UV completion of our model (the full string compactification in the brane picture). Thus, before taking the predictions seriously, one should investigate whether there exists a more complete theory including gravity for which the coefficient of $H^2 \phi^2$ is either naturally small or can be fine-tuned to an acceptable level.
\vskip 0.1 in
\noindent
{\bf The end of inflation}
\vskip 0.1 in
\noindent
A particularly nice feature of our mechanism is that inflation ends naturally -- with no remaining light degrees of freedom in the inflationary sector -- without having to invoke any additional fields, antibranes, etc. In the string theory scenario, this occurs since the brane self-annihilates when it reaches the IR where the cycle on which the brane is wrapped contracts to a point. In the field theory scenario, the theory at the origin of moduli space is a confining theory, in which the confinement scale can be far below the scale of inflation.\footnote{In this case, the theory will be heated well into its deconfined phase when inflation ends. In the string theory scenario, the scale of inflation is generally not high enough to deconfine the gauge theory dual to the throat geometry. For details see \appref{app-reheat}.}  It is conceivable that the non-supersymmetric confining gauge theory left over at the end of inflation could be part the Standard Model (e.g. the Yang-Mills part of QCD), or perhaps a separate sector that gives rise to the observed dark matter.
\vskip 0.1 in
\noindent
{\bf Relation to previous work}
\vskip 0.1 in
\noindent
There exist many previous discussions of inflation models based on supersymmetric gauge theory (see \cite{Lyth:1998xn} for an early review) and of string theory models based on branes in warped throat geometries (for reviews of inflation in the context of string theory, see \cite{Baumann:2009ni, HenryTye:2006uv, Burgess:2007pz, McAllister:2007bg}), but none employ the basic mechanism we have described. From a field theory point of view, the work \cite{Inami:2009bs} is somewhat related in that it uses a higher-dimensional gauge theory with a supersymmetry-breaking compactification, but their candidate inflaton comes from the Wilson line of the higher-dimensional gauge field around the circle. In string theory, models based on wrapped branes have been discussed before \cite{Becker:2007ui}, but here the wrapped branes already have a (relatively steep) potential in a supersymmetric throat geometry. The well-known brane-antibrane inflation model \cite{Kachru:2003sx} is similar in that the inflaton field is the radial position of a brane in a warped throat, and the potential is generated only after supersymmetry breaking. In that case, the supersymmetry breaking occurs by the addition of an anti-brane at the tip of the throat, and inflation ends when the brane and antibrane annihilate. Our model is slightly simpler in that the only brane required is the one that gives rise to the inflaton. Another rather novel feature of our model relative to models studied previously is that we are able to carry out the analysis both in a weakly coupled limit, where we can use field theory methods, and at strong coupling, where we exploit the gravitational dual descriptions.
\vskip 0.1 in
\noindent
{\bf Outline}
\vskip 0.1 in
\noindent
The outline of the paper is as follows. We begin in \secref{sec-setup} by describing the field theory on which our model is based, and giving the gravity dual descriptions that may be used to analyze the theory at strong coupling. In \secref{sec-actions}, we derive the effective action for our candidate inflaton field. We obtain the effective action by directly integrating out W-bosons at one loop and by a dual gravity calculation at strong coupling. The effective actions are in general not of canonical form, but we show in \secref{sec-pot} that during any possible period of inflation, the theory is well-approximated by an effective action with canonical kinetic terms and a simple potential, which we derive in each limit. In \secref{sec-analysis}, we write down the conditions that arise from demanding that slow-roll parameters are small, and derive the predictions for inflationary perturbations in terms of the parameters of our model, assuming a minimal coupling to gravity. In \secref{sec-numbers}, we compare these predictions with observations, thereby constraining the model parameters. We find that our theory gives a viable model of inflation consistent with observations for a broad range of parameters, and summarize the final predictions as a function of the parameters once all observational constraints have been taken into account. In \secref{sec-reheat}, we make some comments about the end of inflation for the two regimes of our model, and discuss a few scenarios for how this model may be integrated with the degrees of freedom of the Standard Model in a more complete theory. In \secref{sec-eta}, we recall that a consistent  UV completion of our field theory may include terms beyond the minimal coupling to gravity that we have assumed, and comment on the potential ``$\eta$-problem'' in which some of these terms may spoil slow-roll inflation. In \secref{sec-related}, we discuss various ways in which our model may be generalized. In \secref{sec-disc}, we make some concluding remarks and discuss future directions.

\section{Basic Setup} \label{sec-setup}

The field theory we use for our main example is maximally supersymmetric 4+1 dimensional Yang-Mills theory compactified to 3+1 dimensions on a circle with antiperiodic boundary conditions for the fermions. This is the theory describing the low-energy physics of a stack of D4-branes (compactified on a circle with twisted boundary conditions) in type IIA string theory. This model was originally introduced by Witten as a construction of pure 3+1 dimensional Yang Mills theory via compactification of maximally supersymmetric 4+1 dimensional gauge theory \cite{Witten:1998zw}. We will summarize the relevant aspects of the construction here.

To define a UV complete theory, we start with a 5+1 dimensional theory, the $(0,2)$ conformal field theory describing the low-energy physics of M5-branes in M-theory. We begin by compactifying this on a circle of size $2 \pi R_{M}$ with supersymmetry-preserving boundary conditions. This gives a theory in 4+1 dimensions whose IR physics is described by the 4+1 dimensional maximally supersymmetric Yang-Mills theory.  We further compactify the theory on a circle of size $2 \pi R$ with antiperiodic boundary conditions for fermions. This theory flows in the IR to a non-supersymmetric confining gauge theory in  3+1 dimensions. The final theory has two dimensionless parameters: a dimensionless coupling
\be
\lambda = {2 \pi N R_{M} \over R}
\ee
which is the 't Hooft coupling in the 3+1 dimensional gauge theory at the Kaluza-Klein scale $1/R$, and the parameter $N$, the rank of the gauge group (which comes from the number of branes). Once the theory is coupled to gravity, the ratio $M_{KK}/M_p = 1/(R M_p)$ will provide a third dimensionless parameter.

\subsection{Weak coupling: $\lambda \ll 1$}

When $\lambda$ is small, the theory is weakly coupled at the Kaluza-Klein scale $M_{KK} = 1/R$ (the much larger scale $1/R_M$ where the 6D physics becomes important will not be relevant in this case). The action is obtained by dimensional reduction of the $d=10$ ${\cal N}=1$ supersymmetric $U(N)$ Yang-Mills theory and gives
\bea
\label{sym}
S &=& {1 \over g_5^2}\int d^5 x \left\{-{1 \over 4} \tr(F_{\mu \nu} F^{\mu \nu}) -{1 \over 2} D_\mu X^I D^\mu X^I + {1 \over 4} \tr([X^I, X^J]^2) \right. \\
 && \qquad \qquad \qquad \qquad \qquad \qquad \qquad   \qquad \left. - {i \over 2} \bar{\psi} \Gamma^\mu D_\mu \psi - {1 \over 2} \bar{\psi} \Gamma^{I+4} [X^I, \psi] \right\} ~, \nonumber
\eea
where the scalars $X^I$ ($I=1,..,5$) and fermions $\psi$ transform in the adjoint representation of the gauge group, and we use ten-dimensional notation for the spinors and gamma matrices. The dimensionful coupling $g_5$ is related to $R_M$ by $g_5^2 = (2 \pi)^2 R_M$, and the dimensionless four-dimensional coupling $\lambda$ is related to $g_5$ by $\lambda = g_5^2 N /(2 \pi R)$.

\begin{figure}[t]
\begin{center}
\includegraphics[width=0.9\textwidth]{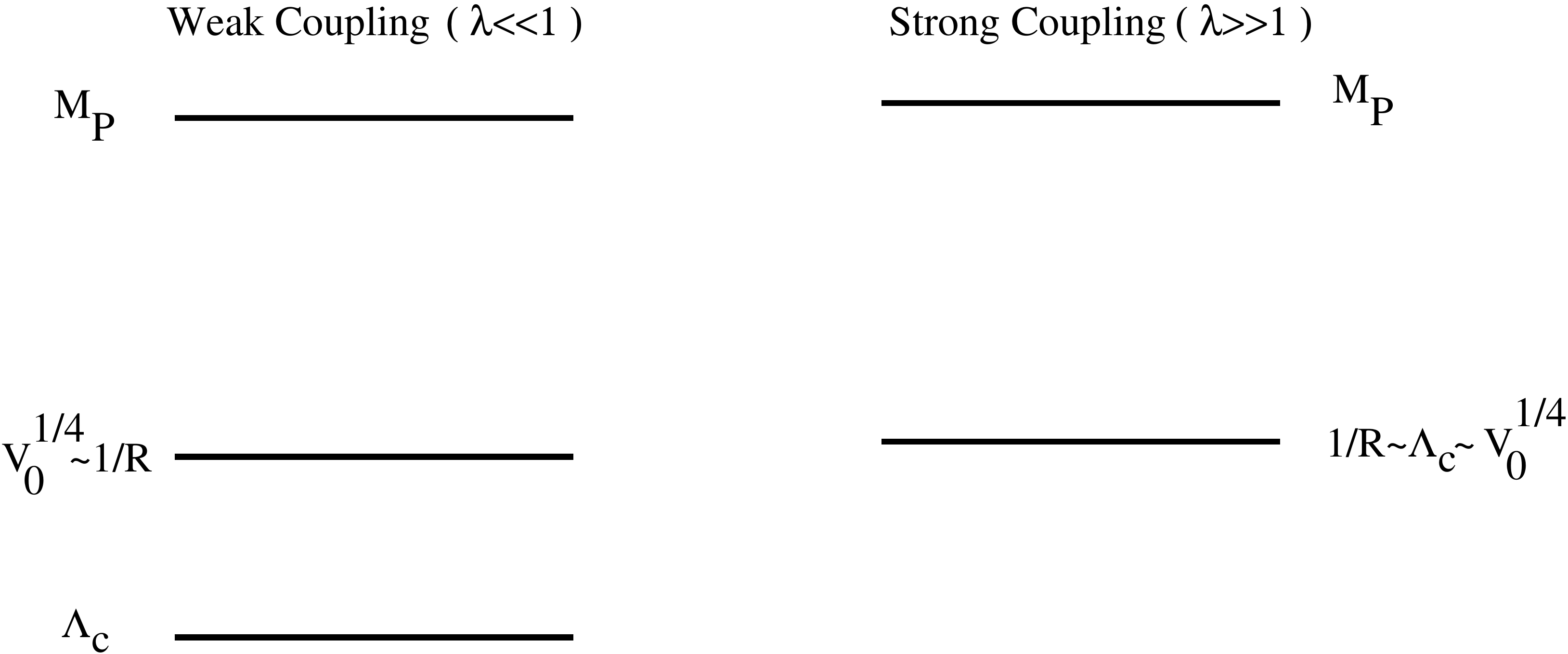}
\end{center}
\caption{\label{scales} Energy scales for weak and strong coupling. The energy density $V_0$ during inflation will be of order the Kaluza-Klein scale $1/R$ in both regimes.}
\end{figure}

As argued by Witten \cite{Witten:1998zw}, the theory flows to pure Yang-Mills theory in the IR with a confinement scale
\be
\Lambda_{conf} \sim {1 \over R} e^{-{c \over  \lambda}} \;
\ee
for some constant $c$ of order 1. At low energies, we have the physics of glueballs with mass of order $\Lambda_{conf}$. The relevant energy scales in this regime are summarized in left hand side on \figref{scales}.

\subsection{Strong coupling: $1 \ll \lambda \ll N^{2 \over 3}$}

For $\lambda \gg 1$, the theory is strongly coupled at all scales, but we can analyze the theory using a gravitational dual description.\footnote{Geometries dual to branes compactified on a circle with antiperiodic boundary conditions for fermions may be obtained by double analytic continuation of the corresponding black p-brane solutions \cite{Itzhaki:1998dd}. } In the regime $1 \ll \lambda \ll N^{2 \over 3}$, the physics is well described by type IIA supergravity on a background \cite{Itzhaki:1998dd}
\bea
\label{IIAthroat}
ds^2 &=& \left({U \over R_4} \right)^{3 \over 2}(\eta_{\mu \nu} dx^{\mu} dx^{\nu} + f(U) dx_4^2) + \left({R_4 \over U} \right)^{3 \over 2}({1 \over f(U)} dU^2 + U^2 d \Omega_4^2)  \cr
e^{\varphi}  &=& \left({U \over R_4} \right)^{3 \over 4} \cr
F_4 &=& 3 R_4^3 \epsilon_4 \; .
\label{metric}
\eea
where $\epsilon_4$ is the volume form on the unit $S^4$, and
\be
\label{ef}
f(U) = 1 - \left({U_0 \over U} \right)^3 \; .
\ee
The $x_4$ direction, corresponding to the KK-direction in the field theory, is taken to be periodic, with coordinate periodicity $2 \pi R$. It is important to note that this $x_4$ circle is contractible in the bulk, so the $x_4$ and $U$ directions form a cigar-type geometry. For this circle to contract smoothly requires
\be
\label{iiaconstants}
R^2 = {4 R_4^3 \over 9 U_0} \; .
\ee
The property that the space ends smoothly at $U=U_0$ is the gravitational signature that the dual gauge theory is confining \cite{Witten:1998zw}.

The parameters appearing in the metric are related to the field theory parameters by
\be
\label{d4coupling}
N = {R_4^3 \over \pi l_s^3} ~, \qquad \qquad \qquad {\lambda \over N} = {2 \pi l_s \over R} \; .
\ee
where $l_s = \sqrt{\alpha'}$ is the string length.

For our purposes it will be useful to encode the Ramond-Ramond flux in terms of the Poincar\'{e} dual five-form potential, $\star F_4 = dC_5$, with
\be\label{iiaform}
C_{5} = \lt({U\over R_4}\rt)^{3}f(U)dx_0 \wedge dx_1 \wedge dx_2 \wedge dx_3  \wedge dx_4~.
\ee
The gauge in (\ref{iiaform}) has been chosen so that $C_5$ vanishes at the tip of the geometry.

In terms of the field theory parameters, the dilaton and the string-frame radius of curvature at the tip of the cigar (the IR part of the geometry) are of order $\lambda^{3 \over 2}/N$ and $\sqrt{\lambda} \ell_s$ respectively. As claimed, supergravity is valid (i.e. the string coupling and the curvature in string units will be small) in the IR region of the geometry when $1 \ll \lambda \ll N^{2 \over 3}$. Note that the string coupling still becomes large in the UV, reaching $\mathcal{O}(1)$ at $U/U_0 \sim N^{4 \over 3}/\lambda^2$. However, for an inflation scenario, we imagine that we have only some finite region of the throat $U < U_{max}$ as part of a consistent compact manifold.\footnote{The limit where the throat becomes infinite corresponds to taking the compactification volume to infinity, which sends the four-dimensional Planck mass to infinity and decouples gravity.} In order to trust calculations based on type IIA supergravity, we require that $U_{max}/U_0 < N^{4 \over 3}/\lambda^2$.

\subsection{Very strong coupling: $\lambda \gtrsim N^{2 \over 3}$}

In the case where $\lambda \gtrsim N^{2 \over 3}$, the string coupling in the type IIA throat becomes large even in the IR part of the geometry, so type IIA supergravity no longer gives a valid description. However, in this regime, the geometry has a weakly curved eleven-dimensional geometrical description, as long as $N$ is assumed to be large.

In this description, the eleven-dimensional metric (obtained from a double analytic continuation of the thermal near-horizon M5-brane solution) is \cite{Klebanov:2000me}
\be
\label{Mthroat}
ds^2 = {\rho \over L}(\eta_{\mu \nu}dx^\mu dx^\nu + dx_M^2 + f(\rho)dx_4^2) + {L^2 \over \rho^2} f^{-1}(\rho) d\rho^2 + L^2 d \Omega_4^2~,
\ee
where
\be
f(\rho) = 1 - {\rho_0^3 \over \rho^3} \; .
\ee
We take the $x_4$ and $x_M$ coordinates to be periodic with periodicities $2 \pi R$ and $2 \pi R_M$ respectively. The parameters in the metric are defined as
\be
L^9 = N^3 {\kappa_{11}^2 \over 2^7 \pi^5}~, \qquad \qquad \rho_0 = {4 L^3 \over 9 R^2}\; ,
\ee
where $\kappa_{11}$ is the eleven-dimensional gravitational coupling. As above, the coupling $\lambda$ in the field theory is related to the metric parameters as
\be
\lambda = {2 \pi N R_{M} \over R} \; .
\ee
The metric is sourced by a four-form field strength $F_4 = 3 L^3 \epsilon_4$, where $\epsilon_4$ is the volume form on the sphere. Again, it is useful for our purposes to describe the flux in terms of the Poincar\'{e} dual potential, $\star F_4 = d A_6$, which gives
\be \label{a6}
A_6 = \lt({\rho\over L}\rt)^{3}f(\rho)dx_0 \wedge dx_1 \wedge dx_2 \wedge dx_3  \wedge dx_4\wedge dx_M~.
\ee

\section{Effective action for the candidate inflaton} \label{sec-actions}

Had we defined our theory with periodic boundary conditions for fermions, the resulting 3+1 dimensional theory would have been supersymmetric, and the theory would have had a moduli space (a family of vacua with zero energy) parameterized by mutually commuting scalar matrices. In the picture where our field theory is understood as the low-energy physics of a stack of $N$ D-branes, these scalar field configurations correspond to configurations of D-branes that are parallel but not coincident: the eigenvalues of the commuting scalar matrices correspond to the coordinates of the branes in the transverse space.

\begin{figure}[t]
\begin{center}
\includegraphics[width= \textwidth]{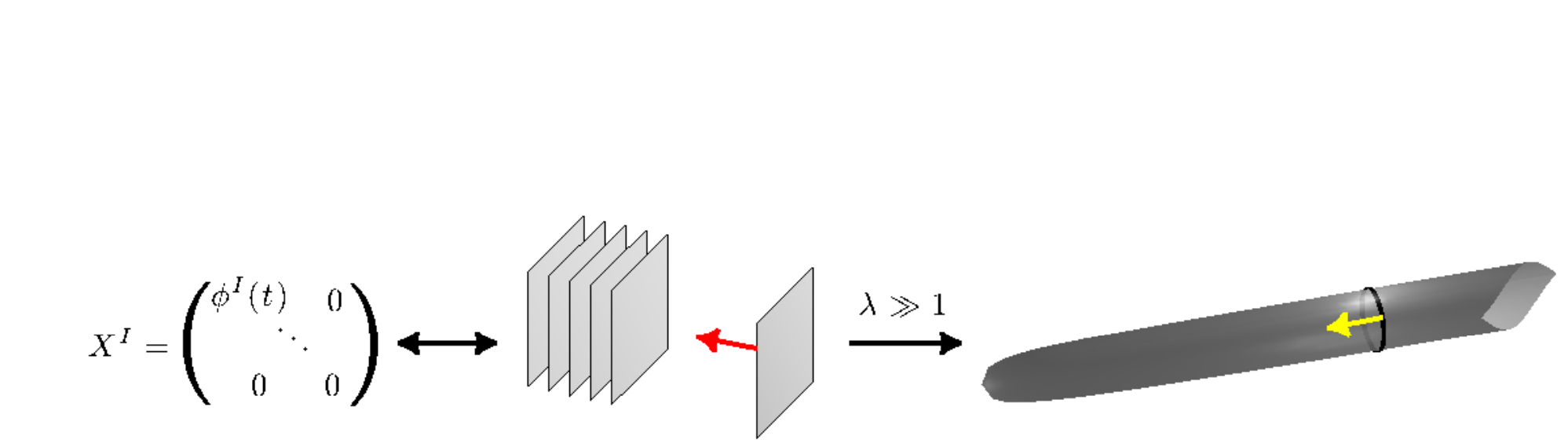}
\end{center}
\caption{\label{inflaton} Various interpretations for the inflaton field.}
\end{figure}

With twisted (antiperiodic) boundary conditions for fermions, the classical potential in (\ref{sym}) still vanishes whenever the scalar fields are described by a set of commuting matrices, but since supersymmetry is broken we will find that a non-trivial potential is generated in the quantum effective action. However, when the scalar field expectation values are significantly greater than the scale of supersymmetry breaking -- the Kaluza-Klein scale -- this potential becomes quite flat.

In this paper, we will mostly consider the simplest nontrivial configuration $X^I = diag(\phi^I,0,\dots,0)$, which in the brane picture corresponds to separating off a single brane from the rest. The absence of supersymmetry results in a net attractive force which tends to bring the isolated brane back to the stack. In this section we will calculate the quantum effective action for the field $\phi^I$. For small $\lambda$, we can do this by a direct field theory calculation (assuming that $|\phi| > M_{KK}$). For large $\lambda$, the physics is described by considering a single brane as a probe rolling down towards the tip of the geometry (\ref{IIAthroat}) or (\ref{Mthroat}), which describe the physics of the remaining branes in the stack (see figure \figref{inflaton}). In all of these regimes, we will show that the potential governing the motion on this lifted branch is naturally rather flat for $|\phi| \gg M_{KK}$, making $|\phi|$ a good candidate for an inflaton field.

\subsection{Weak coupling}

In this section, we consider the gauge theory (\ref{sym}) in the case of small $\lambda$, expanded around a background $X^I = diag(\phi^I,0,\dots,0)$ and derive the effective action for the inflaton field, $\phi\sim|\phi^I|$ by a direct field theory calculation. We begin with the action (\ref{sym}), and expand the theory about a background
\be
X^I = diag(\phi^I,0,\dots,0) \; .
\ee
To make the formulae simpler, we consider the $U(N+1)$ theory. We find that all the fields have off-diagonal modes (the $(1,n)$ and $(n,1)$ matrix elements with $2 \le n \le N + 1$) with masses of order $|\phi|$. These are the W-bosons (together with their scalar and fermionic partners) associated with the breaking of gauge symmetry $U(N+1) \to U(N) \times U(1)$. During inflation, we will assume $|\phi|$ to be initially much larger than the Kaluza-Klein scale $1/R$ (which is roughly the scale at which inflation occurs), so it makes sense to integrate these out to find an effective action for $\phi^I$.

As we show in appendix A, for $|\phi| \gg 1/R$, the kinetic term for $\phi^I$ is dominated by the canonical one coming from the tree-level action, so we can focus on calculating the effective potential (i.e. non-derivative terms), which has no tree-level contribution. A properly gauge-fixed treatment is discussed in appendix A but the result for the effective potential can be reasoned out without lengthy calculation. The one-loop correction to the potential is given as usual by a trace over the logarithm of the inverse propagator
\be
\label{potform}
V^{(1)}(\phi) ={1 \over 2} \int {d^4 k\over (2\pi)^4}\lt\{\sum_{bosons} \log \lt(k^2 +M_b^2\rt) - \sum_{fermions} \log \lt(k^2 +M_f^2\rt)\rt\}~,
\ee
where the sum is over all real field modes being integrated out. In the 5D picture, we have $N$ complex W-bosons with three polarizations, $5N$ complex scalars, and $8N$ complex fermions, all with mass $\phi^2 = \phi^I \phi^I$. In the 4D picture, each of these leads to a KK-tower of particles, with masses
\bea
M_b &=& \sqrt{\phi^2 + {n^2 \over R^2}}~, \cr
M_f &=& \sqrt{\phi^2 + {(n+ {1 \over 2})^2 \over R^2}}~.
\eea
The masses are different because the antiperiodic boundary conditions imply that the allowed momenta in the compact direction are different for bosons and fermions. Thus, while we start with a supersymmetric theory in five dimensions, the potential (\ref{potform}) will be non-zero when we compactify to four dimensions with twisted boundary conditions.

Evaluating (\ref{potform}) in our case, we find that the one-loop effective potential is\footnote{In string theory language, the coefficient here simply counts the number of fundamental string excitations which stretch from the stack of $N$ $D4$-branes to the single brane which has been separated. Each such string is characterized by the member of the stack on which it ends as well as its ten-dimensional polarization, of which there are eight (complex) physical modes for both bosons and fermions. There are thus $8N$ such strings.}
\be\label{vqft}
V(\phi) =8N \sum_{n\in \mathbb{Z}} \int {d^4 k\over (2\pi)^4}\lt\{ \log \lt(k^2 +\phi^2 + {n^2 \over R^2}\rt) - \log \lt(k^2 +\phi^2 + {(n+ {1 \over 2})^2 \over R^2}\rt)\rt\}~.
\ee
This gives a good approximation to the full effective potential, since the tree-level potential for $\phi$ vanishes and higher-loop contributions are suppressed by powers of $\lambda$.

The integral in (\ref{vqft}) is divergent but the finite piece can be isolated by differentiation
\be
{\partial V_{eff} \over \partial \phi^2} =  8 N \int {d^4 k \over (2 \pi)^4} {2 \pi R \over \sqrt{\phi^2 + k^2} \sinh(2 \pi R \sqrt{k^2 + \phi^2})}~,
\ee
where we have used
\bea
\sum_{n\in \mathbb{Z}} {1 \over n^2 + q^2} &=& {\pi \over q} \coth(\pi q)~, \cr
\sum_{n\in \mathbb{Z}} {1 \over (n + {1 \over 2})^2 + q^2} &=& {\pi \over q} \tanh(\pi q) \; .
\eea
Integrating, we find that\footnote{Here, we are assuming that the vacuum energy $V_{eff}(\phi=0)$ vanishes. In reality, this should be equal to the present day vacuum energy, but this would have a negligible effect during inflation.}
\be
V_{eff}(\phi) = {2N C_W \over \pi^2 (2 \pi R)^4} W(2 \pi \phi R)~,
\ee
where we define the function
\be
W(y) = {1 \over {C_W}} \int_0^y dx x \int_0^\infty {q^3 dq \over \sqrt{q^2 + x^2} \sinh(\sqrt{q^2+x^2})} \; ,
\ee
and constant
\be
C_W = \int_0^\infty dx x \int_0^\infty {q^3 dq \over \sqrt{q^2 + x^2} \sinh(\sqrt{q^2+x^2})} = {93 \over 8} \zeta(5) \approx 12.054285\; ,
\ee
so that $W(\infty) = 1$. For large $y$, the function $W$ behaves as
\be
W(y) = 1 - {4 \over C_W}(y^2 + 3y + 3)e^{-y} + {\cal O}(e^{-2y})~.
\ee
Thus, the leading behavior of the effective potential for large $\phi$ is
\be
V (\phi)= V_0\left(1- A \left({\phi \over \phi_0} \right)^2 e^{-{\phi \over \phi_0}} \right)~,
\ee
where
\bea\label{ftparams2}
V_0 &=& {2 N C_W \over \pi^2 (2 \pi R)^4}~,\cr
\phi_0 &=& {1 \over 2 \pi R}~,\cr
A &=& {4 \over C_W}~.
\eea
Since we show in \appref{app-ft} that $\phi$ is canonically normalized, the effective action at weak coupling is
\be
S =  - \int d^4 x \left( {1 \over 2} (\partial_\mu \phi )^2 + V(\phi) \right) \, .
\ee

\subsection{Strong coupling: $1 \ll \lambda \ll N^{2 \over 3}$, the D4-brane} \label{sec-strong}

For large $\lambda$, a more geometrical language is appropriate in describing the dynamics of the inflaton field. The degrees of freedom in the unbroken part of the gauge theory are well described by gravitational physics in the geometry (\ref{IIAthroat}).
The remaining relevant light degree of freedom, corresponding to the scalar field $\phi$,  is described by adding a probe $D4$-brane \cite{Karch:2002sh} in the geometry of (\ref{IIAthroat}), extended along the $x^\mu$ directions and wrapped on the direction $x_4$. In a static gauge, the probe brane's worldvolume is parameterized by the spacetime coordinates $x^\mu$ and $x_4$. We consider an ansatz where the brane is localized on the $S^4$ and has a radial profile $U(x^\mu)$. Note that we assume that $U$ is independent of the twist circle $x_4$. The induced metric on the probe is given by
\be\label{iiainduced}
ds^2 =\left({U \over R_4} \right)^{3 \over 2}\lt(\lt(\eta_{\mu \nu} + \left({R_4 \over U} \right)^3 {\d_\mu U \d_\nu U \over f(U)}\rt)dx^{\mu} dx^{\nu} + f(U) dx_4^2\rt)~.
\ee
The Ramond-Ramond form has a non-trivial pull-back which is (\ref{iiaform}) with $U$ now a function of $x^\mu$.

The action for a single probe $D4$-brane is given by the abelian Born-Infeld action
\be
\label{dbi}
S = - \mu_4 \int d^{5} \sigma e^{-\varphi} \sqrt{-\det(g_{ab} + \tilde{F}_{ab})}~,
\ee
together with the Chern-Simons term
\be
\label{CS}
S = \mu_4 \int \sum_p C_p \wedge e^{\tilde{F}}~,
\ee
where
\be
\tilde{F} = 2 \pi l_s^2 F \; ,
\ee
and
\be
\mu_4 = {1 \over (2 \pi)^4 l_s^5}~.
\ee

For the scenario at hand, in which we do not excite the worldvolume gauge field, the only relevant terms in the brane action are
\be
S = -\mu_4 \int d^{4}x \, dx_4 e^{-\varphi} \sqrt{-\det(g_{ab})} + \mu_4 \int C_5~.
\ee
Evaluating this action for the induced metric (\ref{iiainduced}) and form field (\ref{iiaform}) yields
\be
\label{probeact}
S = (2V_0) \int d^{4} x \left\{-z^3 \sqrt{1-z^{-3}}\sqrt{1 + {9 R^2 \over 4} {1 \over z^3 -1} (\partial_\mu z)^2} + z^3 - 1 \right\}~,
\ee
where we have introduced the dimensionless coordinate $z=U/U_0$ and the constant
\be \label{v0}
V_0 = {2 \lambda N \over 3^6 \pi^2 R^4}~.
\ee
We have expressed the final action entirely in terms of the field theory parameters $N, \lambda,$ and $R$, having made use of (\ref{iiaconstants}) and (\ref{d4coupling}).

\subsection{Very strong coupling: $\lambda \gtrsim N^{2 \over 3}$, the M5-brane} \label{sec-verystrong}

In this section we show the same effective action is valid in the M-theory regime. Here, the rolling scalar corresponds to the radial location of a single M5-brane wrapped on the $x_4$ and the $x_M$ directions in the geometry (\ref{Mthroat}).

Assuming that the worldvolume two-form field is set to zero, the action for a single M5-brane is
\be
S = -T_5 \int \sqrt{-det(g_{ab})} + T_5 \int A_6~,
\ee
where $A_6$ is the six-form field given in \eqref{a6}. Here, the five-brane tension is
\be
T_5 = {1 \over (2 \pi)^5} M_{11}^6~,
\ee
where the eleven-dimensional Planck mass\footnote{We use the conventions of \cite{Polchinski:1998rr}.} is related to $\kappa_{11}$ by
\be
2 \kappa_{11}^2 = (2 \pi)^8 M_{11}^{-9} \; .
\ee
Evaluating the action in the background above, and integrating over the compact $x_4$ and $x_M$ directions, we get
\be
S = {T_5 \over L^3}(2 \pi R)(2 \pi R_M)\rho_0^3 \int d^4 x \left\{-z^3 \sqrt{1 + {L^3 \over \rho_0 z^3} f^{-1}(z) (\partial_\mu z)^2}\sqrt{f(z)} + (z^3 - 1)\right\}~,
\ee
where we have defined $z = \rho/\rho_0$. Expressing everything in terms of the field theory parameters, we find the action (\ref{probeact}). Thus, the effective action (\ref{probeact}) can be considered to be valid for all values $\lambda \gg 1$.

\section{Coupling to gravity and slow-roll potentials} \label{sec-pot}

In the previous section, we computed the effective action for our system assuming a Minkowski space background for the geometry in the four non-compact dimensions and without dynamical gravity (i.e. with $M_p = \infty$ ). To study inflation, we need to include dynamical gravity. In this case, the ratio between the KK-scale and the Planck scale $M_{KK}/M_p = 1/(R M_p)$ becomes a third dimensionless parameter in the model, together with $\lambda$ and $N$.

In the strong coupling picture, studying the field theory without gravity corresponds to considering an infinitely long throat in the dual description (as we did above). In a theory with dynamical gravity, we would have only a finite portion of the throat, appearing as part of some consistent compactification of string theory / M-theory to four dimensions (see \figref{fig-throat}). In this case, the full spacetime  (in the eleven-dimensional picture) is described by a warped metric
\be
ds^2 = w(y^i) g_{\mu \nu}(x) dx^\mu dx^\nu + \tilde{g}_{ij}(y) dy^i dy^j \; ,
\ee
where the $y^i$ are the internal directions, and the four-dimensional Planck mass is given as an integral over the compact directions (including the throat region),
\be
\label{planck}
M_p^2 = {1 \over {\kappa_{11}}^2} \int d^7 y~ w(y^i) \sqrt{\tilde{g}} \; .
\ee
A similar formula holds in the IIA picture.

In this section (and in \secref{sec-analysis} and \secref{sec-numbers}), we will consider the inflationary dynamics of our field theory {\it minimally} coupled to gravity with a standard Einstein action. However, starting with a consistent string compactification (or more generally, a consistent UV complete theory that includes dynamical gravity) the complete effective action will generally also include operators suppressed by powers of the Planck mass, as well as couplings between the fields and spacetime curvature tensors. We postpone a discussion of these corrections, the most dangerous of which can potentially spoil inflation in what is known as the ``$\eta$-problem,'' to section \secref{sec-eta}.\footnote{For the impatient reader, the upshot will be that for a given compactification, there are several different corrections that can give order one contributions to $\eta$, with either sign.  Given that we only require $|\eta | \sim 10^{-2}$ or smaller to obtain slow-roll inflation that agrees with current observations, we would expect that some reasonable fraction of compactifications have this property. More optimistically, there may be a class of compactifications in which this cancelation happens naturally. }

To analyze the behavior of the scalar and find the gravitational backreaction we will make use of a generalized formalism for inflation models with general Lagrangians that are functions of a single scalar field and its first derivative \cite{Chen:2006nt}.  At weak coupling, the scalar field is already canonical and so can be treated by the standard slow-roll analysis. At strong coupling, a more detailed analysis is necessary, and we will show that under a suitable field redefinition we can reduce this model to the canonical slow-roll form, albeit with a different potential than at weak coupling.

We will briefly sketch the generalized formalism; for a more detailed treatment the reader is directed to \cite{Chen:2006nt} and references therein. Consider an action
\be \label{genaction}
S={1\over 2} M_{p}^2 \int d^4 x \sqrt{-g} R + \int d^4x \sqrt{-g} P(X,z) \, ,
\ee
where $z$ is the inflaton field and $X \equiv -{1 \over 2} g^{\mu \nu} \partial_\mu z \partial_\nu z$. Since we are interested in inflationary solutions, we make the ansatz that the metric in the four non-compact directions can be written in Friedmann-Robertson-Walker (FRW) form
\be
ds^2 = -dt^2 +a^2(t) \, dx_3^2 \, ,
\ee
where $t$ is comoving time and we have assumed  spatial flatness. The Einstein and inflaton equations of motion reduce to (a dot indicates a derivative with respect to the time $t$)
\bea
3 M^2 _{p} H^2 &=& E \, , \\
\dot{E} &=& - 3 H (E+P) \, ,
\eea
where $E \equiv 2 X P,_X -P$ is the ``energy'' of the inflaton and $H=\dot{a} / a$ is the Hubble scale factor. The action \eqref{genaction} can give rise to inflation if the ``generalized slow-roll parameters'' (not to be confused with the canonical ones we will define below)
\be
\epsilon_G \equiv  -{\dot{H} \over H^2}  = {3 (E+P) \over 2 E }\, , \qquad \eta_G \equiv  {\dot{\epsilon} \over \epsilon H } \, , \qquad s_G \equiv {\dot{c_s}\over c_s H}
\ee
are all $\ll 1$, where the ``speed of sound'' $c_s$ is defined by
\be
c_s ^2 = {d P \over dE} = {P,_X \over P,_X + 2 X P,_{XX} } \, .
\ee
We will shortly show how to relate these to the usual slow-roll parameters when we are in that regime. We will analyze our system at both weak and strong coupling and show that there are inflationary solutions for all values of $\lambda$.

\subsection{Weak coupling: $\lambda \ll 1$}

At weak coupling we showed in the previous section that the effective action for $\phi R \gg 1$ is
\bea
S&=&  - \int d^4 x \left( {1 \over 2} (\partial_\mu \phi )^2 + V(\phi) \right) \, , \\
V(\phi) &=& V_0 \left(1-A \left( {\phi \over \phi_0} \right)^2 e^{ -{\phi \over \phi_0}} \right) \, ,
\label{Vweak}
\eea
where $V_0$, $A$ and $\phi_0$ are given in \eqref{ftparams2}. We minimally couple this to gravity
\be
S = {1\over 2} M_{p}^2 \int d^4 x \sqrt{-g} R  - \int d^4 x \sqrt{-g} \left( {1 \over 2} (\partial_\mu \phi )^2 + V(\phi) \right) \, .
\ee
This is of the form \eqref{genaction} with $P = X - V $. The inflaton is already canonically normalized, so $V$ is the canonical potential. We find $c_s=1$ and
\be
\epsilon_G = {3 X \over X+V} .
\ee
For this to be small, we require the energy $E=X+V \approx V$, which tells us it is dominated by the potential. Solving the scalar equation of motion with this assumption we find the inflationary attractor solution
\be
\dot{\phi} \approx - {V'(\phi) \over 3 H} \, ,
\ee
where a prime denotes differentiation with respect to $\phi$. Using this solution, we can show that $|\eta_G| \ll1$ if
\be
M_{p} ^2 \left| {V'' \over V} \right| \ll 1\, .
\ee
These conditions are clearly the same as the conditions on the usual slow-roll parameters, and with a canonically normalized scalar and the attractor solution we can relate them
\be
\epsilon = \epsilon_G = {M_{p}^2 \over 2} \left( { V' \over V} \right)^2 \, , \qquad \eta = - {\eta_G \over 2} + 2 \epsilon_G=M_{p}^2 {V'' \over V} \, ,
\ee
where the quantities without subscripts are the usual slow-roll parameters. In this case, the conditions on the generalized parameters guarantee a solution of the canonical ones. In \secref{sec-analysis} and \ref{sec-numbers} we will analyze the potential \eqref{Vweak} in the context of slow-roll inflation.

\subsection{Strong coupling: $\lambda \gg 1 $}

We have shown in \secref{sec-strong} and \ref{sec-verystrong} that both the D4 and M5 probe branes in flat space can be described by the same effective action \eqref{probeact}. The minimal coupling to gravity for these probes is
\be
S={1 \over 2} M_{p}^2 \int d^4x \sqrt{-g} R +V_0\int d^{4} x \sqrt{-g} \left\{-h(z)\sqrt{1 + j(z) (\partial_\mu z)^2} + q(z) \right\} ~,
\ee
where
\bea
h(z) &=& 2 z^3 \sqrt{1- z^{-3} }~, \nonumber \\
q(z) &=& 2 (z^3-1)~, \nonumber \\
j(z) &=& {9 R^2 \over 4} {1 \over z^3-1}~.
\eea
The above action is of the form \eqref{genaction} with
\bea
P &=& V_0 \left( -h(z) \sqrt{1 - 2 j(z) X}+q(z) \right)~, \nonumber \\
E &=& V_0 \left( {h(z) \over \sqrt{1 - 2 j(z) X} }-q(z) \right)~, \nonumber \\
E+P &=&  V_0 \left( {2 h(z) j(z) X \over \sqrt{1-2 j(z) X} } \right)~.
\eea

\begin{figure}[t]
\begin{center}
\includegraphics[width=0.5\textwidth]{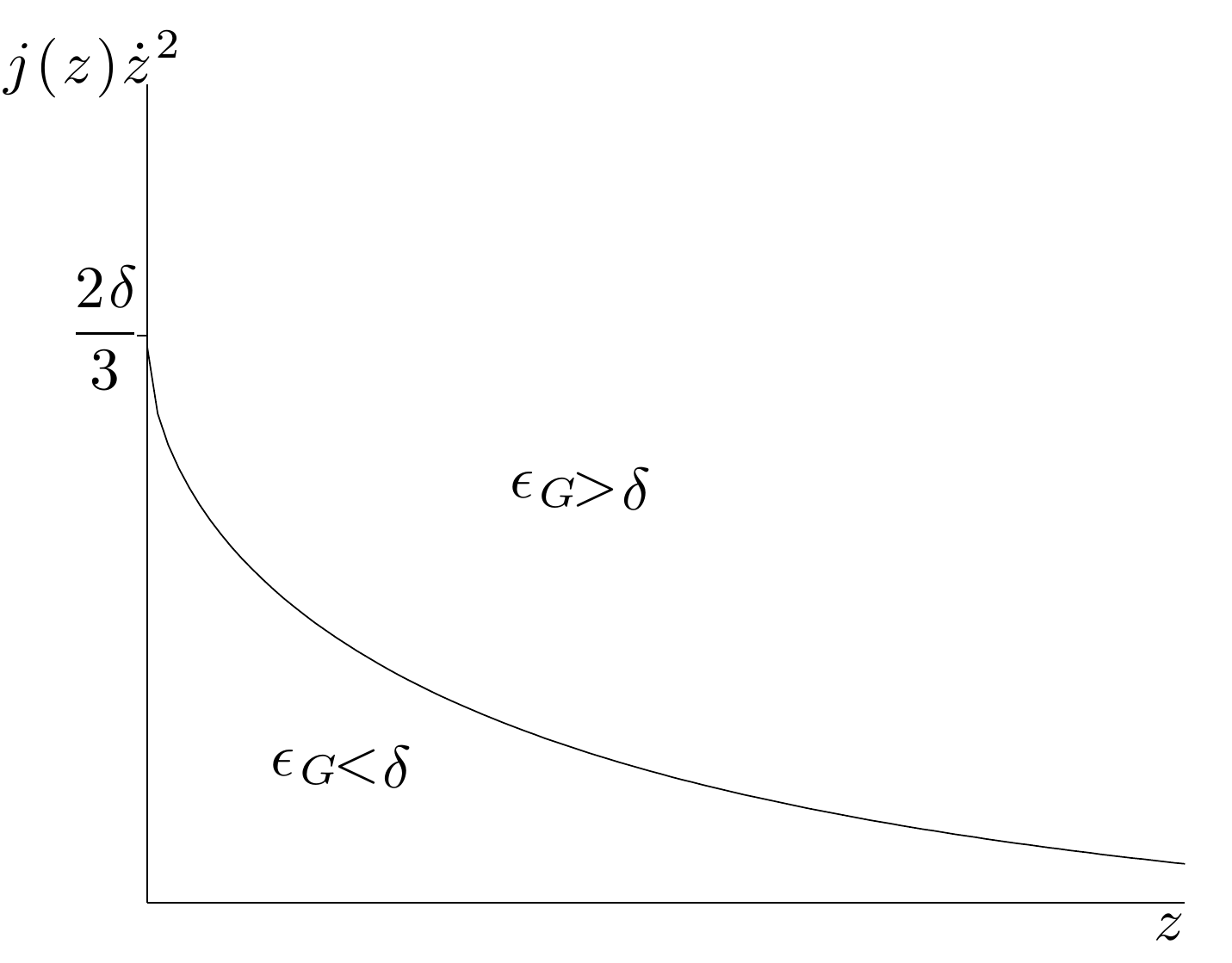}
\end{center}
\caption{\label{epsilon} Diagram showing values of $z$ and $j \dot{z}^2$ for which $\epsilon_G < \delta$.}
\end{figure}

We'd like to search for inflationary solutions in this action. The scalar field $z$ is non-canonical and so we must start with the generalized slow-roll parameters. First, we require
\be
\epsilon_G ={ 3 h(z) j(z) \dot z^2 \over 2( h(z) - q(z) \sqrt{1-j(z) \dot z^2} ) }  \ll 1~.
\ee
We'll now show that whenever this is satisfied, it is a good approximation to keep only the leading terms in the Taylor expansion of the square root. Considering $\epsilon_G$ as a function of $z$ and $j(z) \dot z^2$, it is straightforward to show that in the region where $\epsilon_G < \delta$ (for any $\delta \ll 1$) we have $j \dot{z}^2 < 2 \delta / 3$, as shown in \figref{epsilon}. Thus, the constraint $\epsilon_G \ll 1$ can only be satisfied if $j \dot{z}^2 \ll 1$. This simplifies our analysis considerably, because it is then a good approximation to expand the square root in the action
\be
\sqrt{1- j(z) \dot{z}^2 } \approx 1 -{1\over 2} j(z) \dot{z}^2 ~,
\ee
and we find the action for $z$ minimally coupled to gravity is
\bea
S &\approx& {1 \over 2} M_{p}^2 \int d^4x \sqrt{-g} R+\int d^4 x \sqrt{-g} \left[ -{1 \over 2} G(z) g^{\mu \nu} \d_\mu z \d_\nu z - V(z) \right]~, \\
G(z) &=&  {9 V_0 R^2 \over 2 \sqrt{1-z^{-3} }}~, \nonumber \\
V(z) &=& 2 V_0 \left[ (1-z^3) + z^3 \sqrt{1-z^{-3}} \right] ~. \nonumber
\eea
This action is quadratic in derivatives, but the kinetic term is still not canonical. We can put it in canonical form with a field redefinition
\bea
\phi (z) &=& \sqrt{ 9 V_0 R^2 \over 2} \int_1 ^z \frac{d\tilde z}{(1-  \tilde z^{-3}  )^{1 \over 4} } \approx \sqrt{9 V_0 R^2 \over 2} \left( z + \xi_M -{1 \over 8z^2} + \mathcal{O}(z^{-5} ) \right)~,\nonumber \\
\xi_M &=& -\frac{\Gamma(2/3) \Gamma(3/4)}{\Gamma(5/12)} \approx -0.779936~,
\eea
where the expansion has been carried out at large $z$ since that is the region where we seek a slow-roll solution. This field redefinition gives a canonically normalized scalar field with potential at large $\phi$
\be
V(\phi) = V_0 \left( 1- {\phi_0^3 \over \phi^3} \right)+\mathcal{O}(\phi^{-4})~,
\ee
where $V_0$ is given in \eqref{v0} and
\be
\phi_0 = {1 \over 2^{2\over 3} } \sqrt{ 9 R^2 V_0 \over 2} = \frac{ \sqrt{\lambda N}} {2^{2 \over 3} 9 \pi R} ~.
\ee
Note that $\phi_0$ and $V_0$ here are different from those quantities at weak coupling. We thus have a canonical scalar field with a slow-roll potential, allowing us to carry out the standard analysis. We will do this in \secref{sec-analysis} and \ref{sec-numbers}.

\section{Analyzing the inflationary potentials}\label{sec-analysis}

In the previous section, we have seen that in both the weakly coupled and strongly coupled regimes, the effective action for our candidate inflaton field $\phi$ may be treated as a slow-roll model with canonical kinetic term and potential of the form
\be
V_0(1- f(\phi/\phi_0))
\ee
where $f(x) \to 0$ for $x \gg 1$. The constants $V_0$ and $\phi_0$, and the function $f(x)$ depend on $\lambda$. We have found that for small and large $\lambda$,
\vskip 0.1 in
\begin{center}
\begin{tabular}{|  c | c | c | }
\hline
\qquad & $\lambda \ll 1  $ & $\lambda \gg 1$  \\   \hline  \, & \, & \\
$V_0$ &\large $ \qquad {N C_W \over 8 \pi^6 R^4 } \qquad$ &\large $\qquad   {2 \lambda N \over 3^6
\pi^2 R^4} \qquad $ \\[10pt]
$\phi_0$ &\large $  {1 \over 2 \pi R} $ &\large $  {\sqrt{\lambda N} \over 2^{2\over
3} 9 \pi R}$ \\[10pt]
$f(x)$ &\large $   {4 \over C_W} x^2 e^{-x}$ &\large $  {1 \over x^3}$ \\[10pt]
\hline
\end{tabular}
\end{center}
\vspace{-10 mm}
\be
\label{Vtable}
\vspace{3mm}
\ee
where $C_W = 93 \; \zeta(5)/8 \approx 12.05$. For intermediate values of $\lambda$, these parameters should interpolate from the small $\lambda$ values to the large $\lambda$ values, but we do not have any analytic methods to compute them for general $\lambda$.

In this section we apply a standard slow-roll inflation analysis to determine which values of our model parameters give a viable model of inflation and to investigate the predictions for the inflationary observables.

\subsection{General results for $V(\phi) = V_0(1- f(\phi/\phi_0))$ potentials}

We begin with a general analysis for potentials $V(\phi)$ which approach a constant $V_0$ for $\phi$ much larger than some value $\phi_0$. We define $x = \phi/\phi_0$ and assume that during inflation $f(x) \ll 1$ and $(f'(x))^2 \ll f''(x)$, which will be true for the specific cases we consider.

The standard slow-roll parameters, which must be small to give an acceptable model of inflation, are
\bea
\epsilon &=& {M_p^2 \over 2} \left( {V' \over V} \right)^2 \approx {M_p^2 \over 2 \phi_0^2} (f'(x))^2 \cr
\eta &=& M_p^2 {V'' \over V} \approx {M_p^2 \over \phi_0^2} f''(x)
\eea
Since $(f'(x))^2 \ll f''(x)$ in our case, $\epsilon$ will be small as long as $\eta$ is small.

The equations of motion in the slow-roll regime reduce to
\be
\ddot{\phi} = - \sqrt{3 \over M_p^2} V^{1 \over 2} \dot{\phi} - V'(\phi)
\ee
During slow-roll, we can ignore the acceleration term, so to a good approximation, we have
\be
\label{eom2}
\dot{\phi} =  - \sqrt{M_p^2 \over 3}{V'(\phi) \over V^{1 \over 2}} \; .
\ee
Using this, we can estimate the number of e-foldings during inflation. Starting with the general formula
\be
{\cal N}_e = \int_{t_i}^{t_f} H(t) dt
\ee
we can use the Friedmann equation, the approximation that the energy is dominated by the potential energy during slow-roll, and the equation of motion $(\ref{eom2})$ to derive
\be
{\cal N}_e = \int_{t_i}^{t_f}\sqrt{E \over 3 M_p^2} dt \approx {1 \over M_p^2} \int_{\phi_f}^{\phi_i} {V \over V'} d \phi
\ee
where we integrate between the final and initial values of $\phi$. In our case, this gives the final result
\be
\label{necalc}
{\cal N}_e \approx - {\phi_0^2 \over M_p^2} \int_{x_f}^{x_i} {dx \over f'(x)}
\ee
We must demand that this is at least 60 in order to solve the standard problems of big bang cosmology. For the functions $f(x)$ appearing in our potential for weak and strong coupling, the condition that ${\cal N}_e$ is large enough is essentially the same as the condition that $\eta$ is small.

It will also be useful to relate the value of $x$ at an arbitrary time during inflation to ${\cal N}$, the number of e-foldings before the end of inflation. By the same arguments, we find
\be
\label{ncalc}
{\cal N} \approx - {\phi_0^2 \over M_p^2} \int_{x_f}^{x} {d \tilde{x} \over f'(\tilde{x})} \; .
\ee
We can now use standard slow-roll results to calculate predictions for the inflationary perturbations. The primordial scalar and gravitational wave power spectra are given as
\bea
\Delta_s^2 &=& P^\zeta_k = {1 \over 12 \pi^2 M_p^6} {V^3 \over (V')^2} \approx {1 \over 12 \pi^2 }{V_0 \over M_p^4} {\phi_0^2 \over M_p^2} {1 \over (f'(x_*))^2} \cr
\Delta_t^2 &=& P^h_\zeta = {2 V \over 3 \pi^2 M_p^4} \approx {2 \over 3 \pi^2}{V_0 \over M_p^4}
\eea
from which we can compute the tensor-to-scalar ratio
\be
r \equiv {\Delta_t^2 \over \Delta_s^2} \approx 8 {M_p^2 \over \phi_0^2} (f'(x_*))^2 \; .
\ee
These quantities are to be evaluated at $x_*$, the value of $x$ when the perturbations of the desired scale left the horizon. For the perturbations that lead to the observed CMB anisotropies, this occurred at a number of e-foldings ${\cal N}_{CMB} \sim 60$ before the end of inflation. From (\ref{ncalc}), we obtain the relation
\be
{\cal N}_{CMB} \approx - {\phi_0^2 \over M_p^2} \int^{x_*}_{x_f} {dx \over f'(x)} \; .
\ee
The scale dependence of the power spectra is conveniently parameterized by scalar and tensor spectral indices $n_s$ and $n_t$, and the running parameter $\alpha_s$. Their definitions and their expressions for our class of potentials are
\bea
n_s - 1 \equiv {d \ln P^\zeta_k \over d \ln k } &=& M_p^2 \left(-3 {V'^2 \over V^2} + 2 {V'' \over V} \right) \approx -2 {M_p^2 \over \phi_0^2} f''(x_*)  \\
n_t \equiv {d \ln P^h_k \over d \ln k } &=& -M_p^2 {V'^2 \over V^2}  \approx -{M_p^2 \over \phi_0^2} (f'(x_*))^2 \cr
\alpha_s \equiv {d n_s \over d \ln k} &=& M_p^4 \left(-6 {V'^4 \over V^4} + 8 {V'^2 V'' \over V^3} - 2 {V' V''' \over V^2} \right) \approx -2 {M_p^4 \over \phi_0^4} f'(x_*)f'''(x_*)
\nonumber
\eea
where for $\alpha_s$, we have assumed that $|f'''(x)| \gg |f''(x)f'(x)| \gg |(f'(x))^3|$ as is the case for the specific potentials appearing in our model.

\subsection{Weak and strong coupling results for the inflationary parameters}

Using the results of the previous section, we can specialize to the specific potentials that we obtained for small and large $\lambda$ in our model, as summarized in (\ref{Vtable}). First, the formula (\ref{ncalc}) relating the variable  $x$ to the number of e-foldings before the end of inflation gives for large $x$
\be
\label{Nvsx}
{\cal N}^{\lambda \ll 1} = {C_W \over 16 \pi^2} {1 \over (R M_p)^2} {e^x \over x^2} \qquad \qquad {\cal N}^{\lambda \gg 1} = {\lambda N \over 2^{4 \over 3} 3^5 5 \pi^2} {1 \over (R M_p)^2} x^5
\ee

\subsubsection{Number of e-foldings}

Taking $x$ to be the initial value of $\phi/\phi_0$ in (\ref{Nvsx}) gives ${\cal N}_e$, the total number of e-foldings during inflation. At first sight, it appears that we could obtain an arbitrarily large number of e-foldings by taking the initial value of $x$ to be sufficiently large. However, at least in the strong coupling framework, we have a bound on $x$ since $\phi$ is limited by the length of the throat in our compactification, and the size of the throat sets a lower bound on the 4D Planck mass by equation (\ref{planck}). Specifically, the integral in (\ref{planck}) must be at least as big as the contribution from the throat region, which yields\footnote{Here, we are using the eleven-dimensional metric (\ref{Mthroat}) in the integral (\ref{planck}) and assuming that the throat extends to a maximum value $z_{max}$. Note that this bound implies that the field is restricted to sub-Plankian values, as found quite generally for inflation models based on D3-branes in type IIB throats in \cite{Baumann:2006cd}. As in that case, this bound implies that the production of gravitational waves (tensor modes) during inflation will be small \cite{Lyth:1996im}.}
\be
\phi_{max} < \sqrt{3 \over N} M_p
\ee
or equivalently
\be
x_{max} \le 2^{2 \over 3} 3^{5 \over 2} \pi {R M_p \over \sqrt{\lambda} N} \; .
\ee
Combining this with the strong coupling result in (\ref{Nvsx}), we obtain the constraint
\be
\label{Nelimbig}
\boxed{{\cal N}^{\lambda \gg 1}_e \le {2^2 3^{15 \over 2} \pi^3 \over 5} {(R M_p)^3 \over N^4 \lambda^{3 \over 2}}} \; .
\ee
For the weak coupling model, it is less clear that there should be an upper bound on $x$, though we might demand $\phi \ll M_p$ in order that Planck-suppressed operators may be ignored. Demanding this gives
\be
\label{xmax2}
x_{max} \ll 2 \pi R M_p \; ,
\ee
so (\ref{Nvsx}) gives a constraint
\be
\label{Nelimsmall}
\boxed{{\cal N}_e^{\lambda \ll 1} \le {C_W \over 64 \pi^4} {1 \over (R M_p)^4} e^{2 \pi R M_p}} \; .
\ee

\subsubsection{Inflationary perturbations}

Starting again from (\ref{Nvsx}), but taking ${\cal N}$ to be the number of e-foldings ${\cal N}_{CMB}$ before the end of inflation when the CMB perturbations left the horizon gives the value $x_*$ needed to compute the spectral parameters. Using this, we can express the results for all spectral parameters in terms of the field theory parameters $\lambda$, $N$, and $1/(RM_p) = M_{KK}/M_p$ and the parameter ${\cal N}_{CMB}$ which should be about 60. The results are given in the table below.

\begin{center}

\begin{tabular}{|  c | c | c | }
\hline
\qquad & $\lambda \ll 1  $ & $\lambda \gg 1$  \\   \hline  \, & \, & \\

$\Delta_s^2$ & \large $\quad {C_W \over 24 \pi^6} {N {\cal N}_{CMB}^2 \over (R M_p)^2}\quad$  & \large $\quad {5^{8 \over 5} \over 2^{1 \over 5} 3^5 \pi^{14 \over 5}}{(\lambda N)^{2 \over 5} {\cal N}_{CMB}^{8 \over 5} \over (R M_p)^{14 \over 5} } \quad$ \\
&& \\
$\Delta_t^2$ & \large ${C_W \over 12 \pi^8} {N \over (R M_p)^4}$ & \large ${4 \over 3^7 \pi^4} {\lambda N \over (R M_p)^4}$ \\
&& \cr
$r$ & \large ${24 \over \pi^2} {1 \over (R M_p)^2 {\cal N}_{CMB}^2}$ & \large ${2^{11 \over 5} \over 3^2 5^{8 \over 5} \pi^{6 \over 5}} {(\lambda N)^{3 \over 5} \over ( R M_p)^{6 \over 5} {\cal N}_{CMB}^{8 \over 5}}$ \\
&& \\
$n_s - 1$ & \large $\qquad \qquad -{2 \over {\cal N}_{CMB}}  \qquad \qquad$ &  \large $\qquad \qquad -{8 \over 5 {\cal N}_{CMB}} \qquad \qquad$ \\
&& \\
$n_t$ & \large $- {1 \over 4 \pi^2}{1 \over (R M_p)^2 {\cal N}_{CMB}^2}$ & \large $- {1 \over 2^{4 \over 5} 3^2 5^{8 \over 5} \pi^{6 \over 5}}{(\lambda N)^{3 \over 5} \over (R M_p)^{6 \over 5} {\cal N}_{CMB}^{8 \over 5} }$ \\
&& \\
$\alpha_s$ & \large $-{2 \over {\cal N}_{CMB}^2}$ & \large $-{8 \over 5 {\cal N}_{CMB}^2}$ \\
&&\\
\hline
\end{tabular}
\label{table-test}

\end{center}
We note in particular that $n_s$ and $\alpha_s$ are independent of the parameters of the model for large and small $\lambda$. Since the small and large $\lambda$ values are different, the result for intermediate $\lambda$ must be some nontrivial function of $\lambda$ that approaches these two constants in the limits.

\subsection{Other light fields}

Up to this point we have used the single field formalism to study the inflationary
potentials. However, there are other light fields closely related to the identified
inflaton in both the weak and strong coupling scenarios that need to be considered,
to ensure they do not spoil our analysis.

In the field theory language, we expanded about a background given by $X^I =
diag(\phi^I,0,\dots,0)$, corresponding to separating off a single brane from the
rest in the brane picture. Our inflaton is the magnitude of this separation, $|\phi| =
\sqrt{\phi^I\phi^I}$, but the angular components of $\phi^I$ also correspond to light fields.
In the strong coupling regime, these fields describe the position of the probe brane in the $S^4$ directions.
It is clear that the potential is independent of these angular components, since configurations with different 
values for these are related by a symmetry.

Another light field arises from the Wilson line of the gauge field on the probe brane around the compact direction.
In the field theory language, this is the Wilson line of the $U(1)$ gauge field around the compact direction when the gauge symmetry is broken from $U(N)$ to $U(N-1) \times U(1)$. From the four-dimensional point of view, this is a scalar field, which does not appear in the potential since a constant shift in the $U(1)$ gauge field is a symmetry of the theory.

The initial values of these extra fields, labeling one of them $\theta$ for
illustration, have no effect on the potential, so we have a family of equivalent
inflationary trajectories labeled by different $\theta$.
We could also consider more general trajectories by allowing the other fields to have an initial velocity\footnote{We have checked that assuming zero initial velocity is not a requirement for inflation.} but here, we restrict to the case where only the radial field varies.

In this case we should still check whether the extra light fields contribute to the power spectrum. We have already made slow-roll assumptions on these fields, which allows us to
make use of the machinery for multi-component inflation expounded in Sec. 4 of
\cite{Lyth:1998xn}. In this case, the spectrum is given by
\be
\label{multicomponent}
\Delta_s^2 \propto V \sum_{b\in \textnormal{fields}} \left({{\cal N}_e,_b}\right)^2~,
\ee
where ${\cal N}_e,_b$ stands for the derivative of ${\cal N}$ with respect to
the field indicated by $b$ evaluated at the value of the field when the
perturbations of the desired scale leave the horizon. For these
massless fields we have ${\cal N}_e,_\theta = 0$, since the values of these fields may be changed by a symmetry transformation.
Thus, the only field that participates in the sum is the inflaton and (\ref{multicomponent}) reduces to
the single field result, justifying our analysis above.

It is also important to ask whether bulk moduli can be ignored during inflation. While this question depends on the details of the compactification, a generic expectation is that the masses of scalar fields associated with the bulk moduli should not be lighter than the scale of supersymmetry breaking. Thus, we expect bulk moduli fields to be heavier than $M_{KK} \sim 1/R$. We will see that this is much larger than the scale $H \sim M_{KK}^2/M_p$ that controls thermal/quantum fluctuations during inflation, so the bulk moduli fields can be safely ignored.

\section{Predictions and Observational Constraints}  \label{sec-numbers}

We are now ready to see which parameter values in our model give predictions consistent with current cosmological observations. To begin, we will assume that ${\cal N}_{CMB} = 60$\footnote{The precise value depends on details of the history of the universe between the end of inflation and the well-understood era after reheating. The value ${\cal N}_{CMB} = 60$ is typical for a reheat temperature not far below the scale of inflation, but this can be somewhat higher or lower. For a detailed discussion, see \cite{Liddle:2003as}.}. This immediately leads to the predictions
\begin{center}

\begin{tabular}{|  c | c | }
\hline
$\lambda \ll 1  $ & $\lambda \gg 1$  \\   \hline   \, & \\

 $\qquad n_s = 0.967 \qquad$  &  $\qquad n_s = 0.973 \qquad$ \\
& \\
 $\alpha_s = -0.00055$ &  $\alpha_s = -0.00044$ \\
& \\
\hline
\end{tabular}
\end{center}
without any further assumptions about the parameters $\lambda$, $N$, or $R M_p$. The results for both regimes are consistent with the current best fits including seven-year WMAP data \cite{Komatsu:2010fb},  $n_s = 0.963 \pm 0.012$ and $-0.061 \le \alpha_s \le 0.017$.

The observed magnitude of scalar perturbations can be used to determine the Kaluza-Klein scale $1/(R M_p) = M_{KK}/M_p$ in terms of the field theory parameters $\lambda$ and $N$. To obtain the observed CMB normalization result $\Delta_s^2 = (2.4 \pm 0.2) \times 10^{-9}$ \cite{Bunn:1996py} from the results above, we must have
\begin{center}
\begin{tabular}{|  c | c | }
\hline
$\lambda \ll 1  $ & $\lambda \gg 1$  \\   \hline   \, & \\
\large $\qquad  {1 \over R M_p}  = {3.6 \times 10^{-5} \over \sqrt{ N} } \qquad$  & \large  $\qquad  {1 \over  R M_p}  = {7.5 \times 10^{-4}  \over (\lambda N)^{1 \over 7}}  \qquad$ \\
& \\
\hline
\end{tabular}
\vspace{-10mm}
\be
\label{kkscale}
\vspace{3mm}
\ee
\end{center}
A further constraint comes from demanding that the number of e-foldings is at least 60. In the weakly coupled regime, we see from (\ref{Nelimsmall}) that we can obtain an essentially arbitrary number of e-foldings for any value of $N$ and $\lambda \ll 1$, since the exponent in the constraint on ${\cal N}_e$ is ${\cal O}(10^5)$. For $\lambda \gg 1$, using the value (\ref{kkscale}) for the Kaluza-Klein scale in (\ref{Nelimbig}) gives
\be
{\cal N}^{\lambda \gg 1}_e \le {2.2 \times 10^{14} \over \lambda^{15 \over 14} N^{25 \over 7}} \; .
\ee
Demanding that the right side is at least 60 gives a constraint
\be
\label{Neconstr}
\boxed{\lambda^{3 \over 5} N \lesssim 10^4}
\ee
Even if we take very large $\lambda$ of order $N$ (corresponding to the theory derived from M5-branes on torus with cycles of similar size), we obtain a sufficient number of e-foldings as long as $N$ is less than about $10^3$. Thus, we can obtain a sufficient number of e-foldings for a very broad range of parameters in the model.

Finally, we note that using the results (\ref{kkscale}) for the Kaluza-Klein scale, the predicted tensor-to-scalar ratio in the two regimes is
\begin{center}
\begin{tabular}{|  c | c | }
\hline
$\lambda \ll 1  $ & $\lambda \gg 1$  \\   \hline   \, & \\
 $\qquad r  = 8.6 \times 10^{-12}  N^{-1} \qquad$  &   $\qquad  r  = 2.5 \times 10^{-9}  (\lambda N)^{3 \over 7}  \qquad$ \\
& \\
\hline
\end{tabular}
\end{center}
Taking into account the constraint (\ref{Neconstr}), we find that the tensor-to-scalar ratio is never more than $10^{-6}$. This is certainly consistent with the current experimental bound $r < 0.24$ \cite{Komatsu:2010fb}, but also at a level that is beyond the sensitivity of planned experiments. The smallness of the tensor modes arises due to the relatively low scale for inflation. Explicitly, we have:
\begin{center}
\begin{tabular}{|  c | c | }
\hline
$\lambda \ll 1  $ & $\lambda \gg 1$  \\   \hline   \, & \\
 $\qquad V_0^{1 \over 4} / M_p = 7.2 \times 10^{-6}  N^{-{1 \over 4}} \qquad$  &   $\qquad  V_0^{1 \over 4} / M_p = 9.7 \times 10^{-5}  (\lambda N)^{3 \over 28} \qquad$ \\
& \\
\hline
\end{tabular}
\end{center}
where we recall that $V_0$ gives the energy density during inflation.

\section{The end of inflation and reheating}\label{sec-reheat}

An interesting feature of our model, both for strong and weak coupling, is that the field theory is confining. That is, expanding about $\phi=0$, all physical particles must be gauge-singlets, just as in QCD. In particular, since the inflaton field transforms non-trivially under the action of the gauge group, inflaton particles do not exist as finite energy excitations about the vacuum state (just as there are no free quarks in QCD). This is clear geometrically in the strong coupling picture, since the brane associated with the inflaton field self-annihilates at the end of inflation. In the weakly coupled regime, the IR physics is that of pure Yang-Mills theory, so the relevant excitations are glueballs with mass of order $\Lambda_c$. In either case, there is no leftover light scalar field to worry about at the end of inflation, and the inflaton should efficiently give up its energy to other degrees of freedom. The details of this process, and the implications of our scenario, will be somewhat different in the two regimes, as we know explain.

\subsection{Weak coupling: $\lambda \ll 1$}

In the small $\lambda$ regime, the energy scale during inflation is of order the $KK$-scale, which is larger than the confinement scale by a factor $e^{c/\lambda}$. Thus, the energy of the inflaton field is certainly enough to drive a deconfinement transition in the field theory at the end of inflation.\footnote{If the confinement scale is sufficiently far below the KK-scale, the Hubble temperature during inflation can be larger than the confinement scale, in which case, the field theory will already be deconfined during inflation.} Eventually, we would have a transition back to a confined phase, with glueballs replacing the deconfined plasma.

In a more complete model, our field theory must at least be supplemented with the degrees of freedom of the Standard Model. Here, there are a few interesting scenarios. First, it is possible that the non-abelian gauge fields in the inflation model are actually the same ones that appear in QCD. It is certainly possible that the Standard Model could arise from a higher-dimensional supersymmetric field theory compactified on a circle with antiperiodic boundary conditions (e.g. see \cite{Barbieri:2001dm} and references therein). If this higher dimensional field theory (before the compactification) has a moduli space parameterized by some scalar fields, then one of these scalar fields could become the inflaton in the compactified theory. In this scenario, reheating is particularly simple, since the energy of the inflaton would be dumped directly into Standard Model degrees of freedom.

Another interesting scenario would be that the field theory we have discussed is part of a dark matter sector. In this case, the glueballs we obtain in this sector at the end of inflation would be a dark-matter candidate. There must be some coupling between the degrees of freedom in this sector with the Standard Model degrees of freedom, so that Standard Model particles are also created during reheating. In this scenario, supersymmetry breaking in the Standard Model sector may arise via these couplings from the explicit supersymmetry breaking in this inflaton/dark matter sector.

Finally, we could have a scenario in which the Standard Model is a separate sector from the inflationary sector, but is coupled to this sector in such a way that the glueballs can decay into Standard Model (and dark matter) particles. In this case, no particles from the inflationary sector would be relevant to the current epoch.

It would be interesting to investigate in greater detail the dynamics of reheating in these models, especially in the weak coupling scenario where the inflaton releases its energy directly into Standard Model degrees of freedom. We leave this for future work.

\subsection{Strong coupling: $\lambda \gg 1$}

At large values of $\lambda$ where the gravity picture is valid, the Kaluza-Klein scale is roughly the same as the confinement scale, so it is less clear whether deconfinement in the inflationary sector will occur at the end of inflation. To decide this, we can compare the energy density of the brane during inflation with the energy density of the deconfined plasma just above the deconfinement transition. As we show in appendix B, the latter energy density is
\be
e_{dec} =  {2 \cdot 5 \over 3^7 \pi^2} {N^2 \lambda \over R^4} ~.
\ee
On the other hand, the energy density carried by the inflaton field at the start of inflation is approximately $V_0$, given in (\ref{Vtable}), so we have
\be
e_{i} = {2 \lambda N \over 3^6 \pi^2 R^4} ~.
\ee
Comparing these, we find that
\be
{e_i \over e_{dec}} = {3 \over 5 N} \; .
\ee
Thus, in the range of parameters where a gravity description is valid, the brane will not have enough energy to drive a deconfinement transition at the end of inflation.

In this case, the reheating phase of inflation will more directly involve the production of glueballs. As in the field theory scenario, these could either decay into Standard Model particles, or be left over as a potential dark-matter candidate.\footnote{For a current review of reheating see \cite{Allahverdi:2010xz}.} In this regime, the confinement scale is too high to associate the confining gauge theory used for inflation with QCD.

\section{The $\eta$-problem} \label{sec-eta}

In the previous sections, we introduced a particular field theory, computed the effective action in Minkowski space, minimally coupled this to gravity, and then studied the resulting theory as a potential model of inflation. Ideally, we would have started with a UV complete theory including gravity (perhaps arising from some consistent compactification of string theory), derived the effective action at the scale of inflation and proceeded to analyze that theory. In our minimal treatment, there are two types of terms in the effective action that we may have missed.
\vskip 0.1 in
\noindent
{\bf Curvature Couplings}\label{sec-curv}
\vskip 0.1 in
\noindent
The first type of corrections can be terms coupling the inflaton directly to the spacetime curvature, for example
\be
\label{curv}
-\int d^4 x {\alpha \over 2} {\cal R} \phi^2 \; .
\ee
These terms make no contribution when studying the field theory in Minkowski space, but clearly modify the potential during inflation, where the background spacetime is approximately de Sitter space with positive spacetime curvature. Since ${\cal R}$ is of order $H^2$ during inflation, the term (\ref{curv}) gives an effective mass term with $m^2 \propto \alpha H^2$ for the inflaton field. When $\alpha$ is of order 1, these terms give order one contributions to the slow-roll parameter $\eta$, which therefore must be canceled by other contributions in order to maintain a viable model.

For a field theory described in terms of a gravity dual, such as our model in the strong coupling regime, we can calculate these contributions directly by perturbing the infinite throat metric so that the boundary metric is de Sitter, and then calculating the inflaton effective action by studying the brane effective action in this background. We carry out this calculation for our model in \appref{app-extra}, and find a leading correction term
\be
V_H = {3 \over 4} H^2 \phi^2 \; .
\ee
This gives a contribution to $\eta$ of
\be
\eta_{H} = {3 \over 2} {H^2 M_p^2 \over V_0} \sim {1 \over 2} \; .
\ee
In order to obtain $\eta \ll 1$ in the complete theory, we must therefore have other contributions to $\eta$ that cancel this one.
\vskip 0.1 in
\noindent
{\bf Extra terms from a consistent UV completion with dynamical gravity}
\vskip 0.1 in
\noindent
The terms we have just described appear even if we study the quantum field theory in curved spacetime without adding dynamical gravity (i.e. keeping $M_p = \infty$). The second type of correction arises from terms that must be added to obtain a consistent UV completion with dynamical gravity. As an example, it is well known that in promoting a globally supersymmetric theory to a supergravity theory, it is necessary to include in the potential an overall factor $e^{K/M_p^2}$, where $K$ is the K\"{a}hler potential. This typically includes quadratic terms for all fields, so in a range of fields where the inflaton potential for the globally supersymmetric theory was approximately constant, $V \sim V_0$, the corresponding supergravity theory will include terms $V_0 \phi^2/M_p^2 $ that again give order one contributions to $\eta$.

For field theories dual to gravity on a throat geometry, there is a simple geometrical picture for the origin of these correction terms. Going from a decoupled field theory to a field theory coupled to gravity corresponds to passing from an infinite throat geometry to that of a finite throat which appears as part of some consistent compactification. Near the tip of the throat, the potential for a probe brane should be essentially the same as in the infinite throat case. However, as we go to the top of the throat, the geometry becomes modified relative to the infinite throat, and this results in modifications to the inflaton potential.

At present, finding fully consistent compactifications which include an appropriate throat region is difficult, so we cannot systematically investigate all possible UV completions to determine the possibilities for the corrected inflaton potential. However, as pointed out in \cite{Baumann:2008kq, Baumann:2010sx}, one can at least classify the possible corrections that can occur by starting with the infinite throat geometry and studying all (non-normalizable) perturbations in the UV consistent with the supergravity equations of motion. One can then check to see which of these affect the inflaton potential and classify the possible corrections that might appear in the context of a true compactification.

In the eleven-dimensional picture, our throat geometry in the UV approaches $AdS_7/T^2 \times S^4$, where by $AdS_7/T^2$ we mean $AdS_7$ with two directions compactified. Thus it may be straightforward to carry out this program using previous results (see \cite{Minwalla:1998rp} and references therein) for the perturbation spectrum of eleven-dimensional supergravity on $AdS_7 \times S^4$. One could then check, as was done in \cite{Baumann:2008kq, Baumann:2010sx} for the case of the Klebanov-Strassler throat in IIB string theory, that there exist allowed perturbations which could (for the right compactification) cancel the effect of the curvature couplings discussed above. Assuming such perturbations are possible at all, one might expect that one in a hundred consistent UV completions  would give rise to sufficient cancelation, since we only require $\eta \sim 10^{-2}$. Here, we are assuming a uniform distribution of order one coefficients for the total $H^2 \phi^2$ term. More optimistically, there could be a class of compactifications in which such cancelation occurs naturally.

\section{Related models}\label{sec-related}

In this paper, we have focused on a particular field theory (or particular brane configuration) and a particular initial condition for which a single scalar field eigenvalue begins with a nonzero value. However, there are many possible generalizations making use of the same basic idea.

\subsection{Multiple scalars and N-flation}

First, we could choose a more generic initial condition in which many scalar field eigenvalues are non-zero (or equivalently, in which the $U(N)$ gauge symmetry is broken not just to $U(N-1)\times U(1)$ but to a more generic subgroup). In the supersymmetric case, any collection of mutually commuting scalar matrices $X^i$ gives zero potential, so we expect that any such configuration (with large enough eigenvalues) could be a viable initial condition for inflation. For the case where the number of nonzero scalar eigenvalues is still small compared with $N$, the potential would take the form
\be
\label{Nflate}
V = \sum_i V(\phi_i)
\ee
with terms depending on differences of the scalar field eigenvalues suppressed by ${1 \over N}$. In the strong coupling geometrical picture, this form arises because we have a number of branes independently falling down towards the tip of the same throat geometry. Models with multiple scalars and a potential of the form (\ref{Nflate}) have been dubbed ``N-flation'' \cite{Dimopoulos:2005ac} (for a specific choice of potential see \cite{Liddle:1998jc}), and are generally even more advantageous for inflation, since each scalar feels the Hubble friction associated with the whole collection of inflatons.\footnote{Further analysis would be required to see whether this ``assisted'' inflation effect works in our model beyond leading order.}

In the most generic case where the gauge symmetry is broken to $U(1)^N$, the scalar field potential will be a more complicated function that (\ref{Nflate}) depending on the differences of the eigenvalues. In this case, there will be no geometrical picture in the strongly coupled regime until the end of inflation when gauge symmetry is restored.

\subsection{Other field theories}

A much broader set of generalizations comes by starting with more general higher-dimensional supersymmetric field theories. Taking some higher-dimensional theory with a moduli space and compactifying to four-dimensions so that supersymmetry is broken by boundary conditions, we again expect the potential far from the origin of moduli space to be quite flat. Another example starting with 16 supersymmetries would be the D5-brane field theory (maximally SUSY 5+1 dimensional Yang-Mills theory) compactified on a torus with antiperiodic boundary conditions for fermions along one direction. In the strong coupling picture, this leads to a potential of the form $V(\phi) = V_0(1 - \phi_0^2/\phi^2)$. Starting with less supersymmetry, there are many more possibilities. For example, many ${\cal N}=2$ theories (i.e. eight supercharges) in four dimensions can be defined by compactifying the 5+1 dimensional $(0,2)$ CFT on a Riemann surface (possibly with defects) \cite{Gaiotto:2009gz, Gaiotto:2009we}. These typically have a moduli space of vacua. If we change the boundary conditions around some cycle of the Riemann surface so that supersymmetry is broken, there should be a potential for the scalar fields that previously parameterized the moduli space. For examples with lower supersymmetry, there are fewer examples where the gravity dual to the strongly coupled theory is known, so a detailed analysis may be more difficult.

\subsection{Other boundary conditions}

Finally, we could consider supersymmetry-breaking boundary conditions more general than the anti-periodic boundary conditions for fermions. In general, if the higher-dimensional theory has a global symmetry, we can consider boundary conditions in which the fields around some cycle come back to themselves only up to an element of the global symmetry group. A model of inflation based on this more general type of twisted boundary condition was discussed in \cite{Inami:2009bs}.

\section{Discussion} \label{sec-disc}

In this paper, we have described a new mechanism for slow-roll inflation. To conclude, it is worth emphasizing a few of its special features:
\begin{itemize}
\item The model arises quite naturally with only two ingredients: supersymmetry and extra dimensions. In compactifying to four dimensions, supersymmetry-breaking boundary conditions are no less natural (and perhaps even more generic) than supersymmetry preserving boundary conditions.
\item These essential features are both crucial ingredients of string theory; thus the model arises quite naturally from string-theoretic degrees of freedom
\item As a brane-inflation model, it is very simple since it requires only a single brane on a natural throat geometry; no additional ingredients are required to ensure a graceful exit at the end of inflation.
\item The model can be treated analytically -- and provides a viable model of inflation -- over a very broad range of parameters.
\item There are no problematic light degrees of freedom in the inflationary sector left over at the end of inflation; the energy in the inflaton field is quickly transferred to other degrees of freedom at the end of inflation
\end{itemize}

Of course, as for most models of inflation involving supersymmetry, our scenario may require some modest fine-tuning (of order one part in one hundred) to avoid the $\eta$-problem, as we have discussed in \secref{sec-eta}. An interesting future direction would be to carry out an analysis of allowed perturbations in the UV part of our throat geometry (following \cite{Baumann:2008kq, Baumann:2010sx}) to check that operators able to counteract the effects of the curvature couplings discussed in \secref{sec-curv} can arise in the context of consistent compactifications of string theory. Other interesting future directions include fleshing out some of the reheating scenarios discussed in \secref{sec-reheat}, and exploring some of the other models of section \secref{sec-related}. Finally, it will be interesting to see whether our model continues to satisfy the observational constraints (particularly on $n_s$) once the more accurate data from the Planck satellite becomes available.

\section*{Acknowledgements}
We thank Nima Arkani-Hamed, Subinoy Das, Justin Khoury, Matthew Kleban, Kris Sigurdson, David Tong, Bret Underwood, and especially Shamit Kachru for helpful comments. This work has been supported in part by the Natural Sciences and Engineering Research Council of Canada, the Alfred P. Sloan Foundation, and the Canada Research Chairs programme. TSL is supported by the Institute for Particle Physics.

\appendix
\section{Field theory calculations} \label{app-ft}

In this appendix, we give details of the calculation of the one-loop effective potential for our inflaton field in the regime where the field theory is weakly coupled at the Kaluza-Klein scale. We begin with the action (\ref{sym}), and expand the theory about a background
\be
X^I = B^I = diag(\phi^I,0,\dots,0) \; .
\ee
We find that all the fields have off-diagonal modes with masses of order $|\phi|$, which we assume to be initially much larger than the Kaluza-Klein scale $1/R$. Thus, it makes sense to integrate these out to find an effective action for $\phi^i$. We begin by calculating the effective potential (i.e. the non-derivative terms). For this, we may assume that the background $\phi$ is independent of the spacetime coordinates, which simplifies the calculation considerably.

For the calculation, it is convenient to use a gauge-fixing term
\be
{\cal L}_{fix} = - {1 \over 2} (\partial_\mu A^\mu - i [B^I, X^I])^2 \; .
\ee
in order to cancel cross terms between $A$ and $X$. The corresponding ghost action is
\be
{\cal L}_{ghost} = - \partial_\mu \bar{C} D^\mu C + [B_I, \bar{C}][X^I, C] \; .
\ee

We can now write down the action for the heavy modes $Y$, $\eta$, $W$, and $c$ (each $N$-component vectors) defined by
\bea
X^I = B^I &+& \left( \ba{cc}  0 & Y^I \cr Y^{I \dagger} & 0 \ea \right) ~,  \qquad \psi = \left( \ba{cc} 0 & \eta \cr \eta^{\dagger} & 0 \ea \right) ~,  \\
A_\mu &=& \left( \ba{cc} 0 & W_\mu \cr W_\mu^{\dagger} & 0 \ea \right) ~,  \qquad C  = \left( \ba{cc} 0 & c \cr c^{\dagger} & 0 \ea \right) ~. \nonumber
\eea
It is also convenient to combine the gauge field modes and the scalar modes into a single field $Y_A$ where $Y_\mu \equiv A_\mu$ and $Y_I \equiv Y^I$. Then expanding the action to quadratic order in the heavy fields (and any order in the background fields), we find
\bea
{\cal L}_{bos} &=& - |\partial_\mu Y_A|^2 - \phi^2|Y_A|^2 ~, \cr
{\cal L}_{ferm} &=& - i \bar{\eta} \Gamma^\mu D_\mu \eta + \bar{\eta} \Gamma^{I+4} \phi_I \eta ~, \cr
{\cal L}_{ghost} &=& - |\partial_\mu c|^2 - \phi^2|c|^2 ~.
\eea
The one-loop effective potential may be obtained from functional determinants of the operators defining the quadratic action, and we obtain\footnote{A similar calculation was recently performed in \cite{Martinec:2009ks}.}
\be
V_{eff} =  8 N \int {d^4 k \over (2 \pi)^4} \sum_{n } \left\{ \ln(k^2 + \phi^2 + {n^2 \over R^2})  - \ln(k^2 + \phi^2 + {(n + {1 \over 2})^2 \over R^2}) \right\} \; .
\ee
as we found in section 3.

To calculate the modifications to the kinetic terms we can use the same setup as above, but we can no longer work with a spacetime-independent background field. In this case, it is convenient to write the background field as
\[
X^I = B^I = diag(\phi^I + K^I,0,\dots,0) ~,
\]
where we take $\phi^I$ to be a constant and $K^I$ to include all possible spacetime dependence of the background field.

The action to quadratic order in the heavy fields (and any order in the background fields) is
\bea
{\cal L}_{bos} &=& - |\partial_\mu Y_A|^2 - \phi^2|Y_A|^2 +  {\cal L}_{bos}^{int} ~, \cr
{\cal L}_{ferm} &=& - i \bar{\eta} \Gamma^\mu D_\mu \eta + \bar{\eta} \Gamma^{I+4} \phi_I \eta +  {\cal L}_{ferm}^{int} ~, \cr
{\cal L}_{ghost} &=& - |\partial_\mu c|^2 - \phi^2|c|^2 +  {\cal L}_{ghost}^{int} ~,
\eea
where the ``interaction terms'' involving the background field $K$ are
\bea
{\cal L}_{bos}^{int} &=& - Y_A^\dagger ((2\phi \cdot K + K^2) \delta_{AB} + 2 F_{AB})Y_B ~, \cr
{\cal L}_{ferm}^{int} &=&  \bar{\eta} \Gamma^{I+4} K_I \eta ~, \cr
{\cal L}_{ghost}^{int} &=& -c^\dagger (2 \phi \cdot K + K^2) c ~.
\eea
Here, we define an antisymmetric quantity $F$ as $F_{\mu I} = i \partial_\mu K_I$ with $F_{\mu \nu} = F_{IJ} = 0$. In our calculation, we will find it convenient to define the propagators for the heavy fields using the $K$-independent terms and treat the terms depending on $K$ as vertices (with two legs).

Then the propagators are
\bea
\langle Y_A(x) Y^\dagger_B(0) \rangle &=& {1 \over 2 \pi R} \int {d^4 k \over (2 \pi)^4} \sum_n {e^{i(k \cdot x + {n \over R} x_p)} \over k^2 + {n^2 \over R^2} + \phi^2} \delta_{AB} ~, \cr
\langle \eta(x) \eta^\dagger(0) \rangle &=& {1 \over 2 \pi R} \int {d^4 k \over (2 \pi)^4} \sum_n {e^{i(k \cdot x + {(n + {1 \over 2})\over R} x_p)}(k_\mu \Gamma^\mu + \phi^I \Gamma^{I+4}) \over k^2 + {(n + {1 \over 2})^2 \over R^2} + \phi^2} ~,
 \cr
\langle c(x) c^\dagger (0) \rangle &=& {1 \over 2 \pi R} \int {d^4 k \over (2 \pi)^4} \sum_n {e^{i(k \cdot x + {n \over R} x_p)} \over k^2 + {n^2 \over R^2} + \phi^2}  \; .
\eea

To find the kinetic terms, we do a perturbative expansion using the vertices defined above, keeping only terms that are quadratic in the derivatives of $K$. After some work, we find
\bea
{\cal L}^{1 \; loop}_{kin} = 4 N \int d^4 x (\partial_\mu \phi_I \partial^\mu \phi_I) \int {d^4 k \over (2 \pi)^4} \sum_n \left\{ {1 \over (k^2 + \phi^2 + {n^2 \over R^2})^2}  - {1 \over (k^2 + \phi^2 + {(n + {1 \over 2})^2 \over R^2})^2} \right\} && \nonumber \\
\qquad  - 32 N \int d^4 x (\phi \cdot \partial_\mu \phi) (\phi \cdot \partial_\nu \phi) \int {d^4 k \over (2 \pi)^4} \sum_n \left\{ {k^\mu k^\nu \over (k^2 + \phi^2 + {n^2 \over R^2})^4}  - {k^\mu k^\nu \over (k^2 + \phi^2 + {(n + {1 \over 2})^2 \over R^2})^4} \right\}  && \nonumber
\eea
where we have redefined $\phi + K \to \phi$, so that now $\phi$ is the full spacetime-dependent inflaton field. The sums over $n$ can again be evaluated analytically and we find that the resulting integrands for $k$ are exponentially suppressed for large $k$ and for large $\phi R$. Thus, for large $\phi$, the kinetic term for the scalar field is just the canonical one coming from the tree-level action.

\section{Energy in the deconfined phase at strong coupling} \label{app-reheat}

In this appendix, we calculate the energy density in the deconfined phase of the inflationary field theory at large $\lambda$. This determines the minimum initial energy in the inflaton field necessary to drive a deconfinement transition at the end of inflation.

The deconfined phase of the gauge theory corresponds to a gravity solution (in the M-theory picture) \cite{Klebanov:2000me}
\be
ds^2 = {\rho \over L}(-f(\rho)dt^2 + dx^2) + {L^2 \over \rho^2} f^{-1}(\rho) d\rho^2 + L^2 d \Omega_4^2 ~,
\ee
where
\be
f(\rho) = 1 - {\rho_0^3 \over \rho^3} \; .
\ee
The temperature is related to $\rho_0$ and $L$ by
\be
{1 \over T} = {4 \pi \over 3} {L^{3 \over 2} \over \rho_0^{1 \over 2}} ~.
\ee
The area of the horizon is
\be
A = {8 \pi^2 \over 3} L^4 \left({\rho_0 \over L}\right)^{5 \over 2} V ~,
\ee
where $V$ is the five-dimensional coordinate volume in the $x^i$ directions. Using the Bekenstein formula $S = 2 \pi A/\kappa_{11}^2$, we find that the entropy density in the field theory is given by
\be
{S \over V} = 2\left({2 \over 3}\right)^6 \pi^3 N^3 T^5 \; .
\ee
Using $dE = T dS$, and $V = (2 \pi R) (2 \pi R_M) V_3$, we obtain
\be
{E \over V_3} = {5 \over 3} \left({2 \over 3}\right)^6 \pi^3 N^3 T^6 (2 \pi R) (2 \pi R_M)
\ee
for the three-dimensional energy density of the M5-brane theory on a torus with radii $R$ and $R_M$ in its high-temperature phase. This phase is preferred for \cite{Aharony:2006da}
\be
T > T_c = {1 \over 2 \pi R} \; ,
\ee
and the energy just above this critical temperature is
\be
e_{dec} =  {2 \cdot 5 \over 3^7 \pi^2} {N^2 \lambda \over R^4} ~.
\ee
This is the minimum energy density for the deconfined plasma in the case where it is thermodynamically preferred.

\section{The inflaton potential for quantum field theory in de Sitter space} \label{app-extra}

In this appendix, we calculate the modification to the effective potential for the inflaton field in the strongly coupled regime that arises when we study the field theory in a de Sitter background (as we have to a good approximation during inflation), but with non-dynamical gravity (i.e. $M_p = \infty$). Similar computations in a different background were performed in \cite{Buchel:2003qj,Buchel:2006em}.

From an effective field theory point of view, these corrections arise due to couplings involving the spacetime curvature tensors.

We recall that the field theory in the regime of very strong coupling is dual to M-theory on the background with metric
\be
ds^2 = {\rho \over L}(\eta_{\mu \nu}dx^\mu dx^\nu + dx_M^2 + f(\rho)dx_4^2) + {L^2 \over \rho^2} f^{-1}(\rho) d\rho^2 + L^2 d \Omega_4^2
\ee
and $N$ units of four-form flux through the sphere. Here
\be
\label{deff}
f(\rho) = 1 - {\rho_0^3 \over \rho^3} \; .
\ee
During inflation, the four-dimensional metric is not Minkowski space but rather an FRW-type metric that should be well approximated by de-Sitter space. Thus, the actual geometry in the throat region should be that of a warped de-Sitter compactification rather than a warped Minkowski compactification. We will use the fact that there is a consistent truncation of eleven-dimensional supergravity on a sphere (with four-form flux through the sphere) to seven-dimensional supergravity with a cosmological constant. Thus, we will search for a seven-dimensional solution of the form
\be
ds^2 = {\rho \over L}(-dt^2 + e^{Ht}d \vec{x}^2 + e^{a(\rho)}dx_M^2 + e^{q(\rho)}dx_4^2) + {L^2 \over \rho^2} f^{-1}(\rho) d\rho^2
\ee
satisfying
\be
{\cal R}_{\mu \nu} - {1 \over 2} g_{\mu \nu} {\cal R} = {15 \over 4 L^2} g_{\mu \nu} \; .
\ee
Here, $f$ should go to 1 at large $\rho$ while $q$ and $a$ should go to zero, so that the boundary metric is $dS_4 \times T^2$.
With this ansatz, it turns out that the Einstein equations can be reduced to a set of decoupled ordinary differential equations for the functions $f(\rho)$, $a(\rho)$, and $q(\rho)$. Setting $L=1$ for now, the function $f$ must satisfy
\be
- 2\rho^4ff''+ 2\rho^4(f')^2  -2\rho^3 f f' -18 \rho^3 f' - 36 \rho^2 f + 36 \rho^2 = 9H^2(\rho^2f'+2 \rho f-4\rho) - 9 H^4 \; .
\ee
Given a solution for $f$, we must have
\be
(a+q)' = {6 \over \rho f} - {6 \over \rho} - {f' \over f} + {3 H^2 \over f \rho^2} ~,
\ee
and defining $W = a' - q'$, we must have
\be
{d \over d\rho} \ln(W) = -{1 \over \rho} -{3 \over \rho f} -{3 H^2 \over 2 f \rho^2} ~.
\ee
We can solve these equations perturbatively for large $\rho$, starting with the exact solution where $a=0$, $q = ln(1-\rho_0^3/\rho^3)$ and $f$ is given by (\ref{deff}). We find (restoring dependence on $L$)
\bea
f(\rho) &=& 1 + {9 \over 10} {H^2 L^3 \over \rho} + {9 \over 80} {H^4 L^6 \over \rho^2} - {\rho_0^3 \over \rho^3} - {9 \over 800} {H^6 L^9\over \rho^3}(3 \ln(\rho) - 1) + {\cal O}(\rho^{-4}) ~, \cr
a(\rho)+ q(\rho) &=&  {3H^2 L^3 \over 2\rho} - {9H^4 L^6 \over 20 \rho^2} + \dots \cr
a(\rho) - q(\rho) &=& {\rho_0^3 \over \rho^3} + \dots
\eea
We can use the perturbed metric in the M5-brane action to determine the corrections to the effective potential for the inflaton. We end up with a leading correction
\be
V/V_0 \to V/V_0 + {3 H^2 L^3 \over 2 \rho_0^3} \rho^2 \; .
\ee
In terms of the canonical field $\phi$,
\be
V \to V + {3 \over 4} H^2 \phi^2 \; .
\ee

\bibliographystyle{utphys}

\bibliography{inflation}

\end{document}